\documentclass{jfm_modified_2016}
\usepackage{graphicx}
\usepackage{amsmath, amssymb}
\usepackage[inactive,tightpage]{preview}

\usepackage[normalem]{ulem}
\usepackage{mathtools}

\usepackage{esvect}

\usepackage{enumitem}

\pdfmapfile{+rsfso.map}
\DeclareSymbolFont{rsfso}{U}{rsfso}{m}{n}
\DeclareSymbolFontAlphabet{\mathscr}{rsfso}

\usepackage{pdflscape}
\usepackage{afterpage}
\usepackage{flafter}
\usepackage{rotating}
\usepackage{fancyhdr}
\fancypagestyle{lscape}{%
\fancyhf{} 
\fancyfoot[LE]{}
\fancyfoot[LO] {}
 
}

\usepackage{bm}
\usepackage{microtype}

\usepackage[pdfauthor={Philippe H. Trinh}]{hyperref}
\usepackage[usenames, dvipsnames]{xcolor}
\hypersetup{colorlinks=true,linkcolor=MidnightBlue,citecolor=MidnightBlue}

\usepackage{natbib}
\usepackage{array}
\usepackage{booktabs}
\usepackage{multirow}

\usepackage{caption}
\usepackage{subcaption}
\usepackage{tabularx}
\newcolumntype{Y}{>{\centering\arraybackslash}X}
\usepackage{wrapfig,lipsum}

\usepackage[percent]{overpic}
\usepackage[ruled,linesnumbered]{algorithm2e}
\usepackage{float}

\usepackage{tikz}

\newcommand*{\ep}{\epsilon}

\renewcommand*{\i}{\mathrm{i}}
\newcommand*{\im}{\mathrm{i}}

\newcommand*{\Oh}{\mathcal{O}}
\renewcommand*{\Re}{\operatorname{Re}}
\renewcommand*{\Im}{\operatorname{Im}}

\newcommand*{\FS}{\mathcal{F}}
\newcommand*{\V}{\mathcal{V}}

\newcommand*{\de}{\operatorname{d\!}{}} 
\newcommand{\dd}[2]{\frac{\de#1}{\de#2}}

\usepackage{mathabx}
\usepackage{esint} 
\def\Xint#1{\mathchoice
   {\XXint\displaystyle\textstyle{#1}}%
   {\XXint\textstyle\scriptstyle{#1}}%
   {\XXint\scriptstyle\scriptscriptstyle{#1}}%
   {\XXint\scriptscriptstyle\scriptscriptstyle{#1}}%
   \!\int}
\def\XXint#1#2#3{{\setbox0=\hbox{$#1{#2#3}{\int}$}
     \vcenter{\hbox{$#2#3$}}\kern-.5\wd0}}

\def\YYint#1#2#3{{\setbox0=\hbox{$#1{#2#3}{\int}$}
     \vcenter{\hbox{\scalebox{1}[-1]{$#2#3$}}}\kern-.5\wd0}}

\def\dashint{\Xint-}

\shorttitle{Stokes surfaces for the Kelvin wave}
\shortauthor{Johnson-Llambias, Fitzgerald, and Trinh}

\title{Three-dimensional exponential asymptotics and Stokes surfaces for flows past a submerged point source}

\author{Yyanis Johnson-Llambias,\aff{1} 
 John Fitzgerald,\aff{2} \\
 \and Philippe H. Trinh\aff{1}\corresp{\email{p.trinh@bath.ac.uk}}}

\affiliation{
\aff{1}Department of Mathematical Sciences, University of Bath, Bath BA2 7AY, UK
\aff{2} Mathematical Institute, University of Oxford, Oxford OX2 6GG, UK
}

\pubyear{XXXX}
\volume{YYY}
\pagerange{xxx--xxx}
\date{\today~[Draft]}

\begin{document}
\maketitle

\begin{abstract}
 When studying fluid-body interactions in the low-Froude limit, traditional asymptotic theory predicts a waveless free-surface at every order. This is due to the fact that the waves are
in fact \emph{exponentially small}---that is, beyond all algebraic orders of the Froude number. Solutions containing exponentially small terms exhibit a peculiarity known as the \emph{Stokes phenomenon}, whereby waves can `switch-on' seemingly instantaneously across so-called \emph{Stokes lines}, partitioning the fluid domain into wave-free regions and regions with waves. In three dimensions, the Stokes line concept must extend to what are analogously known as \emph{`Stokes-surfaces'}. This paper is concerned with the archetypal problem of uniform flow over a point source---reminiscent of, but separate to, the famous Kelvin wave problem. 
In theory, there exist Stokes surfaces \emph{i.e.} manifolds in space that divide wave-free regions from regions with waves. Previously, in \cite{lustri-2013} the intersection of the Stokes surface with the free-surface, $z=0$, was found for the case of a linearised point-source obstruction. Here we demonstrate how the Stokes surface can be computed in three-dimensional space, particularly in a manner that can be extended to the case of nonlinear bodies.
\end{abstract}
\section{Introduction}
\noindent In the classic Kelvin wave problem, one considers the production of waves in a uniform stream as flow passes a ship modeled as a point source at the origin. As shown by Kelvin [cf. \cite{darrigol-2005}], the mathematical model can be posed in terms of Fourier integrals, after which an asymptotic analysis in the downstream limit predicts the well-known V-shaped wave pattern [see \emph{e.g.} \cite{eggers-1992}, \cite{pethiyagoda2014apparent}]. In situations where the point source is submerged, the analysis is rendered more difficult \citep{lustri-2013}, and this is a scenario we shall discuss in the paper.  

The Kelvin point-source model provides a geometrical linearisation of more complicated wave-generating bodies; however, where it is important to analyse flows around blunt-bodied obstructions with nonlinear geometries, it can be advantageous to develop asymptotics in the low-Froude or low-speed limit \citep{ogilvie-1968}. This is characterised by small values of the Froude number, $F$, defined via
\begin{equation}
\ep \equiv F^2 = \frac{U^2}{gL},
\end{equation}
which provides a measure of the relative balance between inertial forces, governed by the velocity and length scales, $U$ and $L$, and gravitational forces, governed by the acceleration due to gravity, $g$.

\begin{figure}
  \centering
  \includegraphics[width=.9\linewidth]{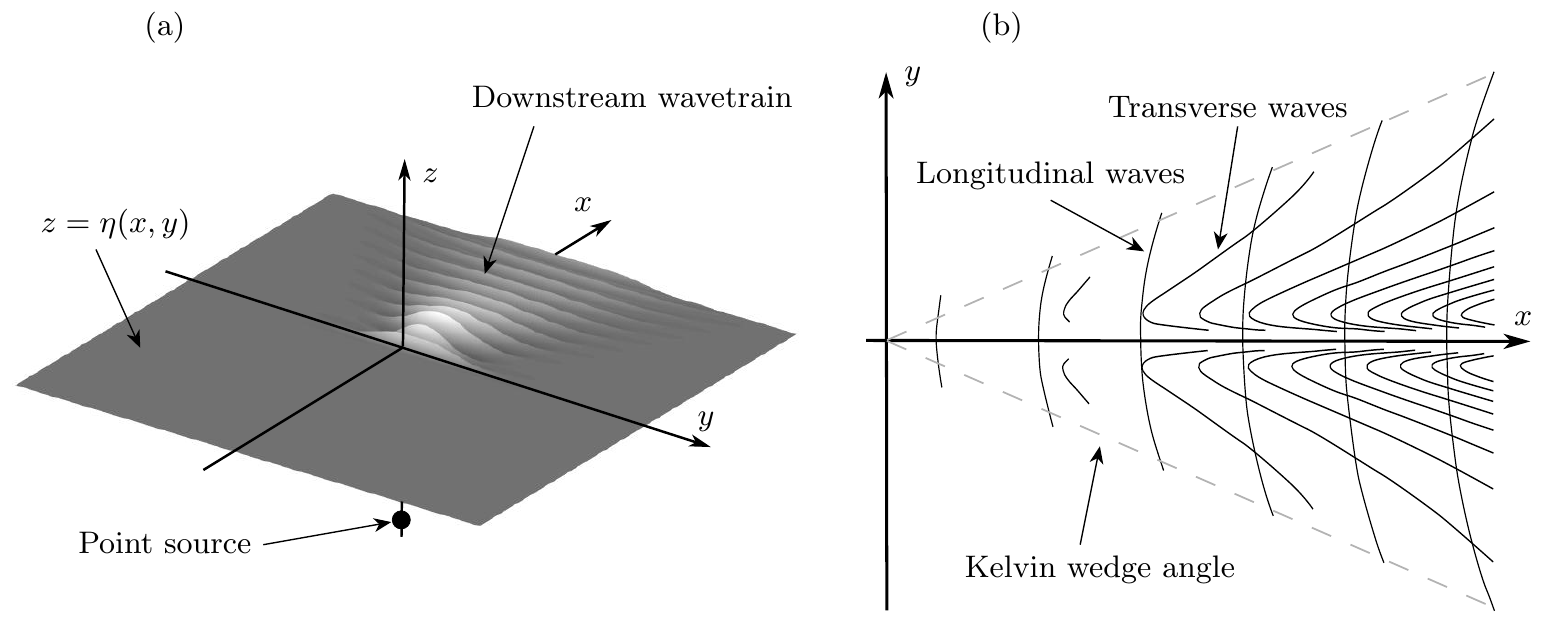}
\caption{\emph{Fluid configuration:} (a) A numerical surface wave calculated for $\ep = 0.15$. The underlying steady flow is in the positive $x$-direction. A point-source (indicated by a bold circle) is placed below the free-surface (indicated by shaded light grey). The interaction of this flow with the point-source induces a downstream wavetrain. (b) The schematic of the wavetrain. Longitudinal and transverse waves form a wavetrain similar to---but distinct from---that seen in the classic Kelvin-ship problem \label{fig:config}}
\end{figure}

Unfortunately, the study of the low-Froude limit, $\ep\to 0$, presents notable challenges first remarked by \cite{ogilvie-1968}, and now referred to as the \emph{`low-speed paradox'} \citep{tulin_2005_reminiscences_and}. Specifically, suppose we write the velocity potential, $\phi$, as a series expansion in $\ep$:
\begin{equation} \label{eq:testphi}
\phi(x,y,z) = \phi_0(x,y,z) + \epsilon \phi_1(x,y,z) + \epsilon^2 \phi_2(x,y,z) +\dots.
\end{equation}
The leading-order term, $\phi_0$, corresponds to the so-called double-body flow where the free-surface surface is flat. This term then encodes the information of the problem geometry; in our case this consists of uniform flow over a submerged point-source at $(0,0,-h)$: 
\begin{equation}\label{eqn:phi0-intro}
\phi_0= \underbrace{Ux}_\text{Uniform flow} -\underbrace{\frac{\delta}{4\pi}\left\{\frac{1}{\sqrt{x^2+y^2+(z-h)^2}}+\frac{1}{\sqrt{x^2+y^2+(z+h)^2}}\right\}}_\text{point-source terms}.
\end{equation}
The leading-order approximation, $\phi_0$, is wave-free, and as remarked by \cite{ogilvie-1968}, the subsequent terms, $\phi_1$, $\phi_2$, etc. will also fail to capture wave phenomena. The waves are in fact exponentially small and beyond-all-orders of the algebraic expansion in $\ep$. 

In fact, as consequence of the singularly perturbed nature of $\ep \to 0$, the base series in \eqref{eq:testphi} diverges. The divergent expansion can be optimally truncated, and the exponentially small remainder sought. This yields
\begin{equation}\label{eqn:phi-trunc}
\phi(x,y,z)=\sum_{n=0}^{N}\ep^n\phi_n(x,y,z) + \Bigl[A(x,y,z)e^{-\chi(x,y,z)/\ep} + \text{c.c.}\Bigr].
\end{equation}
Above, $\chi(x,y,z)$ is an important function known as the \emph{singulant}. We will typically write the real-valued waves in complex exponential form (c.c. for complex conjugate). Thus, $\Re\chi$ provides a measure of the exponential dependence of the waves and this is shown in Fig.~\ref{fig:config}. \emph{Exponential asymptotics} provides those tools for derivation of the exponentially-small contributions \citep{Boyd-1999}.

\begin{figure}
  \centering
  \includegraphics[width=.6\linewidth]{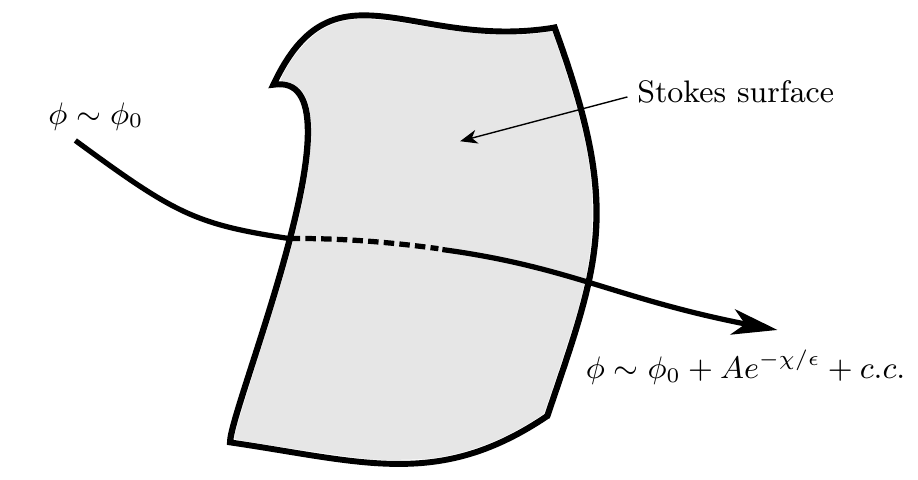}
\caption{\emph{The Stokes surface:} traversing a Stokes surface induces a switch-on of an exponential term in the solution. \label{fig:cartoon}}
\end{figure}

There is a subtle aspect involving the exponentially-small waves in \eqref{eqn:phi-trunc}. These waves do not exist at all points in $\mathbb{R}^3$, but rather, they may switch-on across certain critical manifolds known as \emph{Stokes surfaces}. That is, in certain regions of the physical flow, $A$ in \eqref{eqn:phi-trunc} is identically zero, but $A$ becomes non-zero upon crossing a Stokes surface; this is illustrated in Fig.~\ref{fig:cartoon}. This peculiar transition is known as the Stokes Phenomenon, and is generic to many singularly perturbed problems. In two-dimensional free-surface flows, the Stokes Phenomenon governs the generation of surfaces waves across Stokes lines \citep{chapman-vb-2006}. In standard differential equations theory, the simple analogue that will be familiar to most readers is the transition from exponential decay to oscillatory in the Airy equation \citep{berry_inbook} that occurs across the origin. 

For a given three-dimensional fluid-structure interaction problem in the $\ep \to 0$ limit, the question is whether the Stokes surface(s) can be established. In this paper, we shall demonstrate a numerical methodology that allows Stokes surfaces to be derived for linear geometrical problems (such as the submerged point-source problem), but that can be generalised to nonlinear geometries.  

\section{Background and open questions}

The study of water waves produced by flows past wave-generating bodies is extensive and we refer readers to the literature reviews found in \cite{stoker_book,wehausen_1960,wehausen_1973,tulin_2005_reminiscences_and}. The key distinction of our work is in the analysis of low-speed or low-Froude flows past submerged point sources. The unique exponential smallness of the waves in this regime distinguish the problem from the many other linear water-wave problems (e.g. the classic Kelvin problem) of the past. 

Classically, \cite{keller-1979} studied the Kelvin wave problem of a moving source \emph{on the free-surface}, and showed the solution of the Eikonal equation is obtained by the method of characteristics, establishing the study of ray theory [see \cite{chapman-1998} and others]. In the case of \cite{keller-1979}, the source lies \emph{on} the physical free-surface, and waves are associated with real-valued rays which emanate from the source point. Two natural questions arise: namely how these techniques relate to problems of a submerged body, and how they relate to bodies possessing non-negligible size: so-called bluff-bodied objects. The first of these questions was studied by \cite{lustri-2013}, who demonstrated that in the case of a \emph{submerged} body the analysis is beyond the scope of traditional real-ray methods as used by \cite{keller-1979}. Instead the specialised technique of \emph{complex-ray theory} is required. This study also highlighted the close association between the Stokes phenomenon and the generation of free-surface waves by interactions with submerged solid bodies. Using complex-ray theory, \cite{lustri-2013} showed that the linearised point-source problem admits explicit solutions on the free-surface.\par
 The family of solutions found for the singulant govern the location of the Stokes lines on the free-surface. In essence, these functions are restrictions of the more general three-dimensional singulant to the free-surface, and thus the Stokes lines discovered by \cite{lustri-2013} are the intersections of three-dimensional Stokes surfaces with the free-surface.  In two-dimensional problems, recent exponential asymptotic analysis [by \cite{trinh-chapman-2013a} and others] has shown that the Stokes lines associated with this phenomenon emanate from certain points on solid boundaries (often cusps or corners of the object). In this paper we shall address the following questions:
\begin{enumerate}[label=(\roman*),leftmargin=*, align = left, labelsep=\parindent, topsep=3pt, itemsep=2pt,itemindent=0pt ]
\item What is the nature of the singulant and the associated Stokes surface (the so-called Stokes structure) in three dimensions?
\item In many wave-structure problems of this type which do not use a linear reduction, we are unable to obtain the singulant in exact form. What methods would be appropriate to use when such exact solutions are not available?
\item Would these methods be suitable for investigating the corresponding nonlinear point-source problem?
\end{enumerate} 
  Finally, we note that the reduction of two-dimensional surface flows to a one dimensional boundary integral is a powerful tool and has been the subject of much study recently [see e.g. \cite{vb_book} for boundary integral numerics, and \cite{crew-trinh-2016} for boundary integral analytic continuation]. However, the application of these methods to three-dimensional problems is problematic due to the reliance on results of single variable complex analysis for which the higher dimensional analogue is not clear. Instead, a direct treatment of Laplace's equation for the velocity potential, $\nabla^2\phi = 0$ must be performed.
\section{The necessity of complex-ray theory}\label{sect:necessity}
\noindent As we have discussed, the primary focus of the study shall be the so-called \emph{`singulant'} function, $\chi(\bm{x})$, which characterises exponential terms of the form $A(\bm{x})\exp(-\chi(\bm{x})/\epsilon)$. A significant part of this work relies on working in higher-dimensional complex space (notably $\bm{x}\in\mathbb{C}^3$), so here we shall clarify why the theory must be developed in this space. For clarity, let us introduce the spaces: 
\begin{align}
\text{real fluid volume:} \ \V  &= \{ \bm{x} = (x,y,z)\in\mathbb{R}^3: \ 0 \leq z \leq \eta(x, y) \}, \\ 
\text{real free surface:}\ \FS &= \{ (x,y)\in\mathbb{R}^2, z = \eta(x,y) \}.
\end{align}
In \S\ref{sect:formulation}, we shall see that the singulant is governed by the boundary value problem involving the \emph{Eikonal equation},
\begin{subequations}\label{eqn:problem-intro}
\begin{align}
\chi_{x}^2+\chi_{y}^2+\chi_{z}^2=0,
& \qquad \text{in the fluid, $\bm{x}\in\V$,}\label{eqn:eik-intro}\\
\chi_z=\chi_x^2,
& \qquad \text{on the free-surface, $\bm{x}\in\FS$.}\label{eqn:bc-intro}\\
\chi = 0 & \qquad  \text{ on $x^2 + y^2 + (z\pm h)^2 = 0$}.\label{eqn:ic-intro}
\end{align}
\end{subequations}
 where subscripts denote partial differentiation. Important to note is that the condition \eqref{eqn:bc-intro} is produced by a linearisation about a uniform flow (\emph{i.e.} $\delta\ll0$ in \eqref{eqn:phi0-intro}). As will be outlined in \S\ref{sect:formulation}, the final condition is required for consistency between leading-order and late-order terms of the velocity potential, $\phi$. Notice that $\chi$ is zero at a location related to the two physical source points at $z = \pm h$. 

More specifically, we face the following problems:
\begin{enumerate}[label=(\roman*),leftmargin=*, align = left, labelsep=\parindent, topsep=3pt, itemsep=2pt,itemindent=0pt ]
\item We might attempt to solve \eqref{eqn:eik-intro} and \eqref{eqn:ic-intro}, imposing some behavioural condition near the two source points $(0,0,\pm h)$ and tracing out rays. However, there is no guarantee that such rays would reach $z = 0$, or that if they were to reach, that they would satisfy \eqref{eqn:bc-intro}.

\item Thus we would attempt to substitute \eqref{eqn:bc-intro} into \eqref{eqn:eik-intro} and solve the limited problem of $\chi_x^2 + \chi_y^2 + \chi_x^4 = 0$ on $z = 0$. However, the source condition \eqref{eqn:ic-intro} is now $x^2 + y^2 + h^2 = 0$, which is not satisfied at any real values of $x$ and $y$. Instead, the problem can be solved as a complex-ray problem with rays originating from $x^2 + y^2 + h^2 = 0$ for complex $x, y \in \mathbb{C}$. 
\end{enumerate}
In this paper we therefore leverage complex-ray theory in a non-standard manner.  Our method is comprised of two parts. The first considers a sub-problem on the complexified free-surface (four-dimensions, of which the physical free-surface is a two-dimensional subspace), while the second part extends the complexified free-surface solution into the complexified fluid-domain (six-dimensions, of which the physical fluid-domain is a three-dimensional subspace). By the method of complex-rays, a solution in real-space, say $\chi$, is typically recovered as a real-space restriction of its analytic continuation $\tilde{\chi}$ in complex-space, \emph{i.e.} $\chi = \tilde{\chi}{\big|}_{\mathbb{R}^3}$.
Complex-ray methods therefore often rely on understanding the relationship between real and complex domains. A visualisation of these spaces is presented in Fig.~\ref{fig:big}. Finally, we note the solubility of the problem \eqref{eqn:problem-intro} via Fourier analysis and the method of steepest descents (for details, see App.~\ref{app:fourier}). However, in utilising complex-ray theory we develop a generalisable framework by which we may study fully nonlinear problems. For discussion, see \S\ref{sect:discussion}.
In summary, this paper shall:
\begin{enumerate}[label=(\roman*),leftmargin=*, align = left, labelsep=\parindent, topsep=3pt, itemsep=2pt,itemindent=0pt ]
\item Present a numerical scheme to recover exponentially-small waves via complex-ray theory.
\item Use this scheme to reproduce the results of \cite{lustri-2013}, and extend the results into the inner-fluid domain, revealing the associated three-dimensional \emph{Stokes structure}.
\item Leverage algebraic solutions of the linearised problem to perform far-field asymptotics of the solutions. In particular, we shall confirm certain conjectures about the connection between the submerged source problem and the famous Kelvin problem [see \cite{kelvin-1887}] in both far-field, and shallow-source limits. 
\end{enumerate}

\section{Mathematical formulation}\label{sect:formulation}
{\noindent
We consider a steady, irrotational, incompressible, free-surface gravity flow with surface tension neglected. After suitable nondimensionalisation, the governing equations are formulated in terms of the velocity potential, $\phi$,
\begin{subequations}\label{whole-problem}
\begin{align}
\nabla^2 \phi = \delta(x,y,z+h) 
& \qquad \text{in the fluid,} \label{laplace}
\intertext{with the kinematic and dynamic (Bernoulli) conditions on the free-surface, $z=\eta(x,y)$,} 
\nabla \phi \cdot \bm{n} = 0  & \qquad \text{on $z = \eta(x,y)$,} \label{kinematic}\\
  \frac{\epsilon}{2} (|\nabla \phi|^2-1) + z = 0 
  &\qquad \text{on $z = \eta(x,y)$}, \label{bernoulli}
\end{align}
\end{subequations}
where $\bm{n}$ is the unit outward-pointing normal to the free-surface. The nondimensional parameter, $\epsilon$, denotes the square of the Froude number, and measures the relative balance between inertial and gravitational forces (which are assumed to act in the negative $z$ direction),
\begin{equation}\label{epsilon-def}
\epsilon = \frac{U^2}{gL}.
\end{equation}
As such, the low-Froude limit corresponds to $\epsilon \to 0$.
 We consider the regime in which $0<\delta\ll\ep$, where $\delta$ quantifies a small perturbation to the uniform flow,
$$
\phi = x + \delta \tilde{\phi}.
$$
A balance in \eqref{bernoulli} requires $\eta = \delta \tilde{\eta}$. Under these conditions, our governing equations become
\begin{subequations}
\label{series-sub}
\begin{align}
\nabla^2 \tilde{\phi} = 0,
& \qquad \text{$-\infty<z<0$,} \label{laplace-delta}\\
\left(\tilde{\eta}_x - \tilde{\phi}_z \right)+\delta\left(\tilde{\phi}_x\tilde{\eta}_x+ \tilde{\phi}_y\tilde{\eta}_y\right) = 0,  & \qquad \text{on $z = 0$,} \label{kinematic-delta}\\
  \left(\ep\tilde{\phi}_x+\tilde{\eta}\right) +\delta\frac{\ep}{2}|\nabla\tilde{\phi}|^2=0,
  &\qquad \text{on $z = 0$.} \label{bernoulli-delta}
\end{align}
\end{subequations}
Neglecting terms of $\Oh(\delta)$ and dropping hats, the velocity potential, $\phi$, and the free-surface, $\eta$, are sought via the asymptotic series 
\begin{align}\label{series-phi-eta}
\phi&=\sum_{n=0}^{\infty}\ep^n\phi_n, &   &\eta=\sum_{n=0}^{\infty}\ep^n\eta_n.
\end{align}
Evaluation at $\Oh(\ep^n)$ yields
\begin{subequations}\label{system-n}
\begin{align}
\nabla^2 \phi_n = 0, 
& \qquad \text{$-\infty<z<0$,} \label{laplace-n}\\
{\eta}_{nx} - {\phi}_{nz}=0,  & \qquad \text{on $z = 0$,} \label{kinematic-n}\\
  {\phi}_{(n-1)x}+{\eta}_n = 0, 
  &\qquad \text{on $z = 0$.} \label{bernoulli-n}
\end{align}
\end{subequations}
In the limit $\epsilon\to 0$ Bernoulli's equation \eqref{bernoulli} implies that $\eta_0=0$. This can be understood as gravity dominance forcing a flat free-surface at leading-order. Thus for the point-source problem, we may use the method of images to infer the leading-order velocity potential,
\begin{equation}\label{lo-vel-pot}
\phi_0(x,y,z) = - \frac{1}{4\pi}\left\{\frac{1}{\sqrt{x^2+y^2+(z-h)^2}}+\frac{1}{\sqrt{x^2+y^2+(z+h)^2}}\right\}.
\end{equation}
The key to capturing the exponentially small terms is to study the behaviour of the dominant terms in the limit $n \to \infty$ in order to deduce the nature of the singulant, $\chi$.  

\subsection{Divergence of the asymptotic expansion}\label{sect:divergence}
\noindent The ideas for estimating divergent tails follow from the work of \cite{chapman-1998} (for ODEs), and \cite{chapman-mortimer-2005} (for PDEs). We note that in \eqref{bernoulli} the asymptotic parameter, $\ep$, multiplies the highest derivative, producing a relationship between higher-order series terms and the derivatives of lower-order terms. Thus a singularity present in the lower-order terms in the series will be transmitted and amplified in the power through subsequent terms and cause the expansion to diverge like a factorial-over-power. This suggests the ansatz
 \begin{align}\label{ansatz}
\phi_n&\sim\frac{A(x,y,z)\Gamma(n+\gamma)}{\chi(x,y,z)^{n+\gamma}}, & \text{and} &  &\eta_n&\sim\frac{B(x,y,z)\Gamma(n+\gamma)}{\chi(x,y,z)^{n+\gamma}},
\end{align}
where $\Gamma$ denotes the Gamma function, and $F, \, \gamma$ and $\chi$ are functions that do not depend on $n$. We emphasise that the function $\chi$ is the same singulant appearing in the exponent of \eqref{eqn:phi-trunc}. Its presence as the denominator in the late-order ansatz provides an interpretation for the condition \eqref{eqn:ic-intro}, namely that $\chi=0$ at singularities of the leading-order solution. Important to note when using this ansatz is that we assume that singularities are well-separated; as shown in \cite{trinh-chapman-2015}, in problems with coalescing singularities it is necessary to use a more general exponential-over-power form, however we shall not encounter such problems. Making use of this ansatz in the system \eqref{system-n}, we may derive from the governing equations for the velocity potential those for the singulant,
\begin{subequations}\label{chi-gov}
\begin{align}
\chi_{x}^2+\chi_{y}^2+\chi_{z}^2=0, 
& \qquad \text{$-\infty<z<0$,} \label{chi-eik}\\
-A\chi_{x}+B=0,  & \qquad \text{on $z = 0$,} \label{chi-kin}\\
  B\chi_{x}-A\chi_{z}=0, 
  &\qquad \text{on $z = 0$.} \label{chi-bern}
\end{align}
\end{subequations}
For a non-trivial $\chi$, the latter two equations demand that
\begin{align}\label{eik-solve-cond}
\chi_z=\chi_x^2, & \qquad \text{on $z = 0$.}
\end{align}
Using this condition, the elimination of $\chi_z$ in the Eikonal equation \eqref{chi-eik} yields the governing equation for the singulant, $\chi$, \emph{on the free-surface},
\begin{align}\label{chi-prob}
\chi_x^2+\chi_y^2+\chi_x^4=0, & \qquad \text{on $z = 0$.}
\end{align}
We note that obtaining free-surface solutions for $\chi$ is critical for the recovery of the singulant away from the free-surface as we shall outline in \S\ref{sect:3d}.
  \subsection{Optimal truncation and Stokes line smoothing}
 \noindent Following \cite{trinh-chapman-2013a} [for a comprehensive exposition, see \cite{lustri-2012}], the optimal truncation point, $N$, of a divergent series is typically where successive terms are approximately equal, \emph{i.e.} where
  \begin{equation*}
    \left\lvert\frac{\ep^{N}\phi_{N}}{\ep^{N - 1}\phi_{N - 1}}\right\rvert \sim 1 .
  \end{equation*}
  By substitution of the late-term ansatz, we see that $N \sim |\chi|/\ep$, and so for any fixed point with $\chi \neq 0$, we have that $N \to \infty$ as $\ep \to 0$. As a result, it is precisely the behaviour of these late-order terms that governs the Stokes switching which we wish to determine. Upon truncation, the asymptotic series \eqref{series-phi-eta} become
\begin{align}\label{series-phi-eta}
\phi&=\sum_{n=0}^{N-1}\ep^n\phi_n + R_N, &   &\eta=\sum_{n=0}^{N-1}\ep^n\eta_n+S_N,
\end{align}
where $N$ is sought so as to minimise the remainders $R_N$ and $S_N$. We note that following \cite{lustri-2013}, choosing $N$ according to the above rule suffices for the purposes of this analysis. For a more detailed examination see \cite{trinh-2011, chapman-1998}. Upon substitution of the truncated forms into \eqref{whole-problem} the series terms are eliminated at each order due to \eqref{series-sub}, and we are left with
\begin{subequations}
\begin{align}
\nabla^2 S_N = 0, 
& \qquad \text{$-\infty<z<0$,} \label{remainder-laplace}\\
 R_{Nz}+S_{Nx}=0
  &\qquad \text{on $z = 0$,} \label{remainder-kinematic}\\
\ep R_{Nx}+S_N=-\ep^N\phi_{(N-1)x}
  &\qquad \text{on $z = 0$.} \label{remainder-bernoulli}
\end{align}
\end{subequations}
Following \cite{lustri-2013}, as $\ep\to0$ this system is satisfied by the WKB-like ansatz
\begin{align}\label{rem-exp}
R_N&\sim\Phi(\bm{x}) e^{-\chi(\bm{x})/\ep}, &   S_N&\sim H(\bm{x}) e^{-\chi(\bm{x})/\ep},
\end{align}
where $\chi(\bm{x})$ is one of the solutions to the governing system for the singulant, \eqref{chi-gov}.
This establishes a connection between the exponentially-small remainder and the late-order ansatz \eqref{ansatz} by the presence of $\chi$ in both. The criterion observed in \cite{dingle-1973} states that an exponential term of the form \eqref{rem-exp} (with $\chi = \chi_1$, say) switches-on a further exponential term (with $\chi = \chi_2$, say) when we have 
 \begin{equation}\label{eqn:dingle-general}
\Im(\chi_1)=\Im(\chi_2), \quad \text{and}\quad \Re(\chi_2)\geq \Re(\chi_1).
\end{equation}
The first may be interpreted as an equal phase requirement, and the second a condition of subdominance. Making use of the Dingle criterion, we see that exponentially small ripples of the form \eqref{rem-exp} first emerge when 
\begin{equation}\label{eqn:dingle-2}
\Im(\chi)=0, \quad \text{and}\quad \Re(\chi)\geq 0.
\end{equation}
This is caused by Stokes switching due to the base solution (represented by $\chi_1=0$, due to its algebraicity in $\epsilon$). We note that the singularity providing the largest switching term is that which leads to the smallest value of $|\Im(\chi)|$ on the real axis. We shall assume this is the nearest singularity to the real axis, as is typically (though not always) the case [see \emph{e.g.} \cite{trinh-chapman-vb-2011}].
%
\section{Complex-ray theory for the singulant}\label{sect:complex}
\noindent 
As noted in the previous section, on the free-surface the singulant, $\chi$, is governed by \eqref{chi-prob}, and moreover it is known that the singulant vanishes at singular points, specifically those given by \eqref{eqn:ic-intro}. On the free-surface, this corresponds to the condition that 
\begin{equation}\label{eqn:ic-freesurf}
\chi = 0 \quad \text{on} \quad x^2+y^2+h^2=0.
\end{equation}
 Following \cite{Ockendon-etal-2003} and others, we proceed to seek a solution by use of Charpit's method. To this end, we introduce
\begin{align}
p& = \chi_x, & q& = \chi_y.
\end{align}
\emph{Rays} are defined as parametrised 
solutions to Charpit's equations, in the variable $\tau$,
}
\begin{equation}
\begin{aligned}\label{lin-charpit}
\dd{x}{\tau}&= 2p+4p^3,         &  \dd{y}{\tau}& = 2q, & \dd{\chi}{\tau} &=2p^4,\\
\dd{p}{\tau} &=0,        &  \dd{q}{\tau} &=  0. 
\end{aligned}
\end{equation}
The initial data, at $\tau = 0$, given by \eqref{eqn:ic-freesurf} closes the system. In order for the initial conditions to be in the desired form, given by
\begin{align}
(x,y,p,q,\chi) = (x_0,y_0,p_0,q_0,0) \quad \text{at $\tau=0$},
\end{align}
 we introduce the parametrisation variable, $s$, such that
\begin{equation}\label{x-y-chi-ic}
x_0(s)=s, \qquad y_0(s) = \pm\i\sqrt{s^2+h^2}, \qquad \chi_0(s)=0.
\end{equation}
We emphasise that $s\in\mathbb{C}$ in general (\emph{cf.} discussions in \S\ref{sect:necessity}). The initial conditions $p=p_0(s)$ and $q=q_0(s)$ may be determined by using the Eikonal equation \eqref{chi-prob} and applying the chain rule to ${\text{d}\chi}/{\text{d}s}$. This gives
\begin{align}
p_0^2+q_0^2+p_0^4&=0,\label{p0-q0-1}\\
p_0+\dd{y_0}{s}q_0&=0.\label{p0-q0-2}
\end{align}
Using these two equations we may eliminate $q_0$ (as long as $y_0^\prime\neq 0, \infty$ \emph{i.e.} we have $s\neq 0,\pm\im h$), and this gives
\begin{equation}\label{p0-1}
p_0^2\left(1+\frac{1}{{y_0^\prime}^2}\right)+p_0^4=0.
\end{equation}
Following our assumption of $y_0^\prime\neq 0$, we see from \eqref{p0-q0-2} that $p_0=0$ leads to the trivial solution for $\chi$. Thus excluding this trivial branch, we obtain initial conditions for $p$ and $q$, given by
\begin{equation}\label{p0-q0-branches}
q_0(s) = -\frac{p_0}{y_0^\prime}, \qquad p_0(s) =\pm \i \left(1+\frac{1}{{y_0^\prime}^2}\right)^{1/2}.
\end{equation}
Substitution of \eqref{x-y-chi-ic} then produces
\begin{equation}\label{p0-q0-solns}
p_0 =\pm \frac{h}{s}\quad\text{and}\quad q_0 = \im p_0\frac{\sqrt{h^2+s^2}}{s}.
\end{equation}
We note that there are two branches contained here, and we shall clarify later which branch choices should be considered.  
\subsection{Analytic solutions}\label{sect:algebraic}
\begin{figure}
\centering
\begin{subfigure}{.5\textwidth}
  \centering
  \includegraphics[width=.9\linewidth]{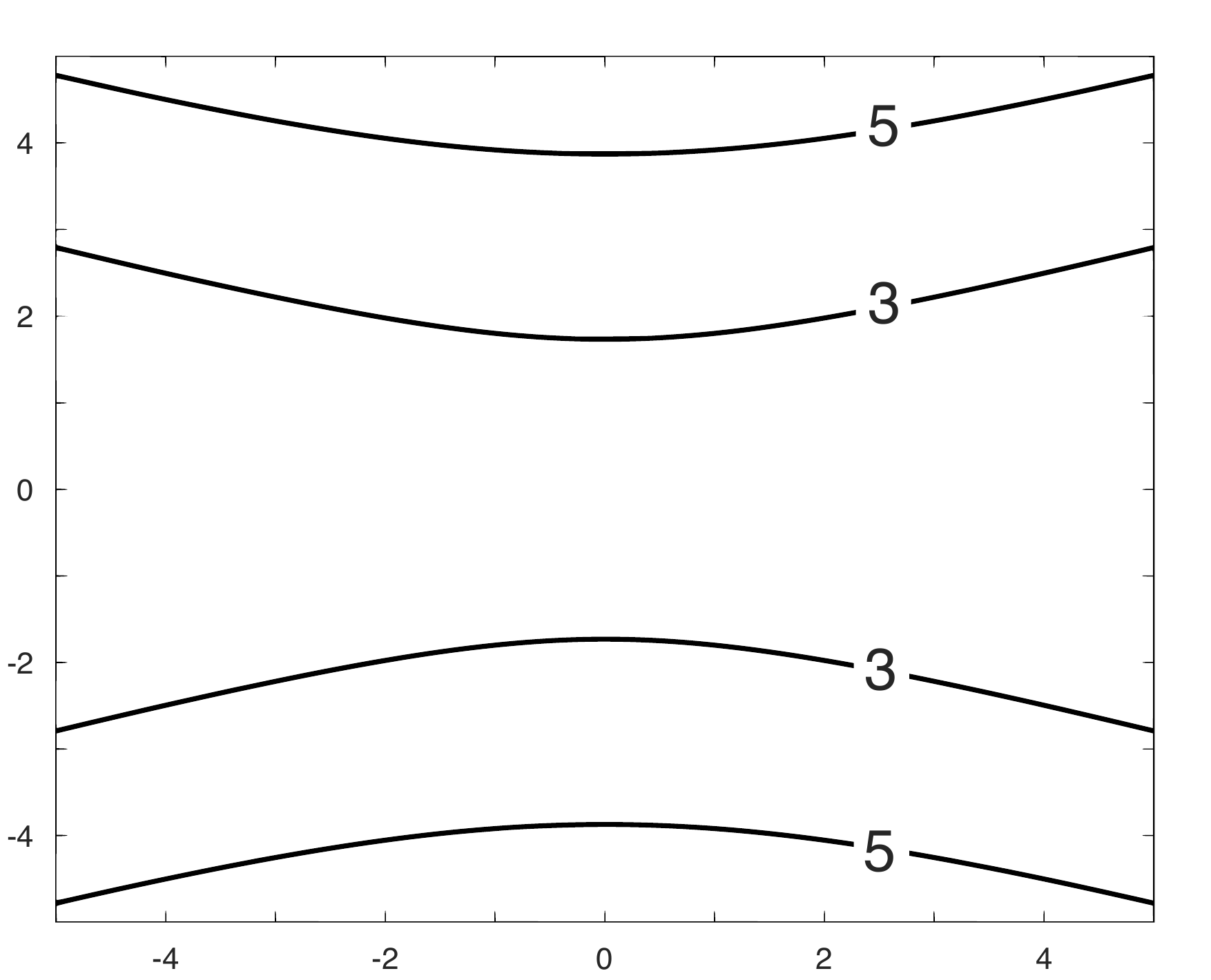}
  \caption{$\Re(\chi_{L1})$}
  \label{fig:bigswalk}
\end{subfigure}%
\begin{subfigure}{.5\textwidth}
  \centering
  \includegraphics[width=.9\linewidth]{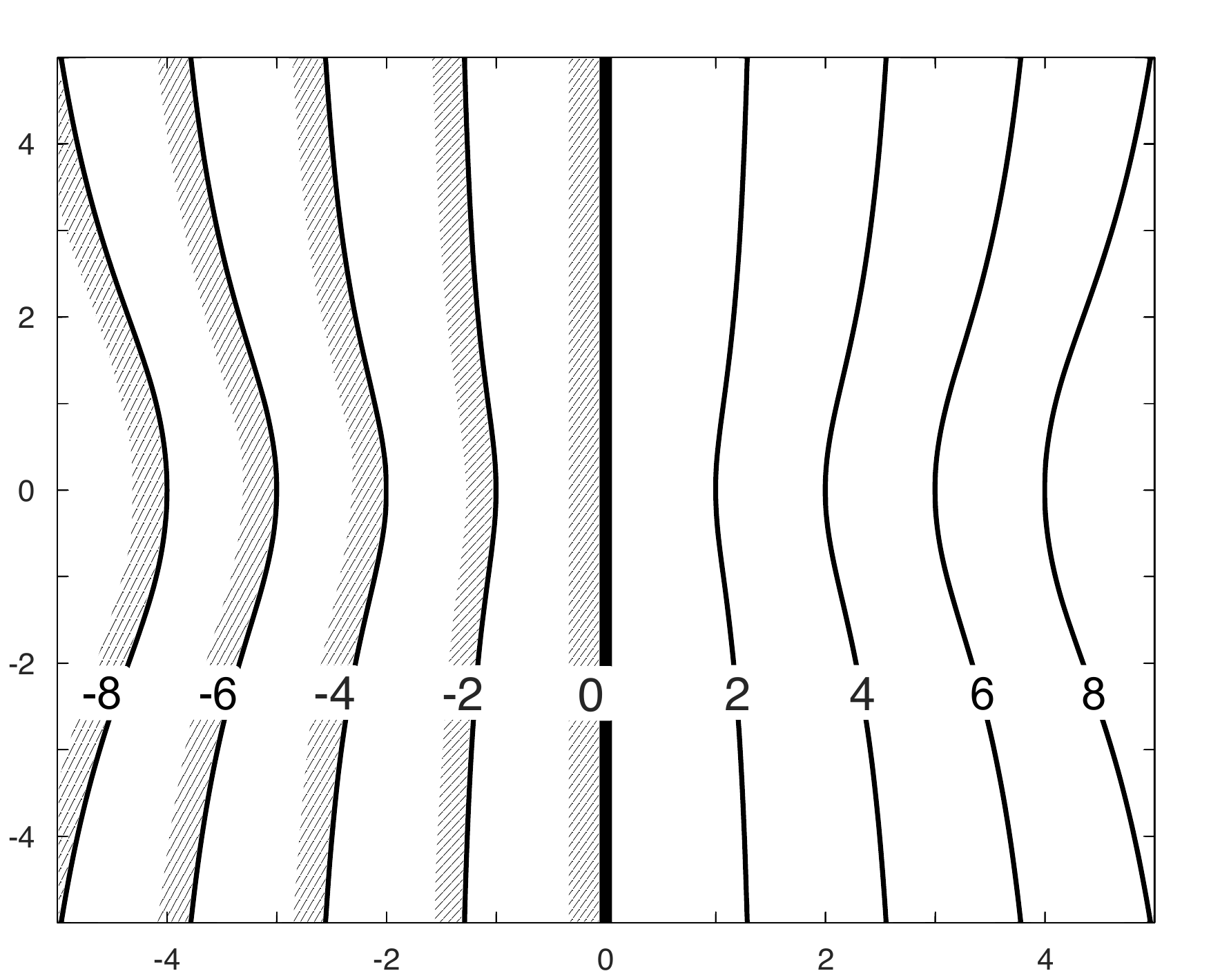}
  \caption{$\Im(\chi_{L1})$}
  \label{fig:bigsxwalk}
\end{subfigure}
\begin{subfigure}{.5\textwidth}
  \centering
  \includegraphics[width=.9\linewidth]{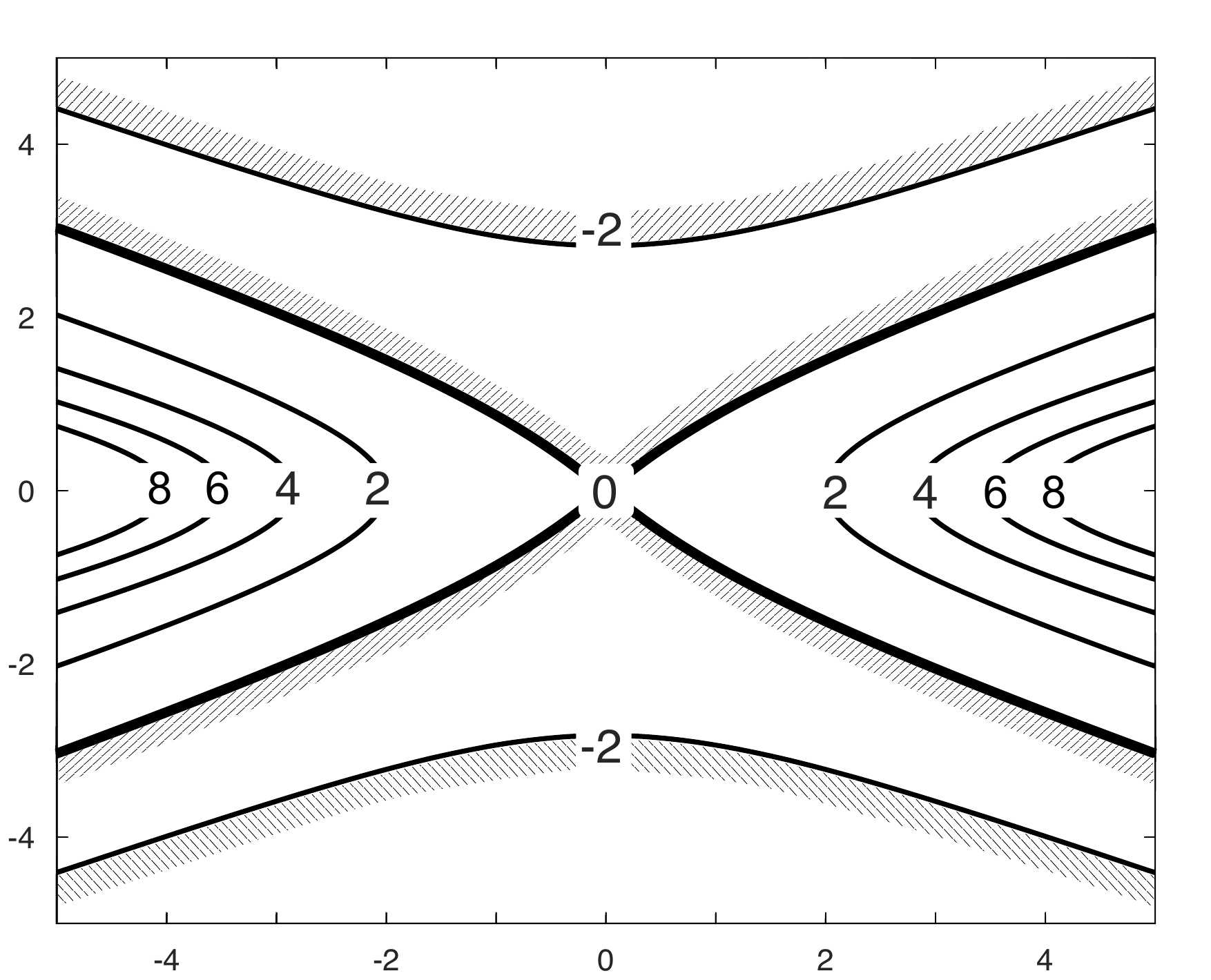}
  \caption{$\Re(\chi_{T1})$}
  \label{fig:bigswalk}
\end{subfigure}%
\begin{subfigure}{.5\textwidth}
  \centering
  \includegraphics[width=.9\linewidth]{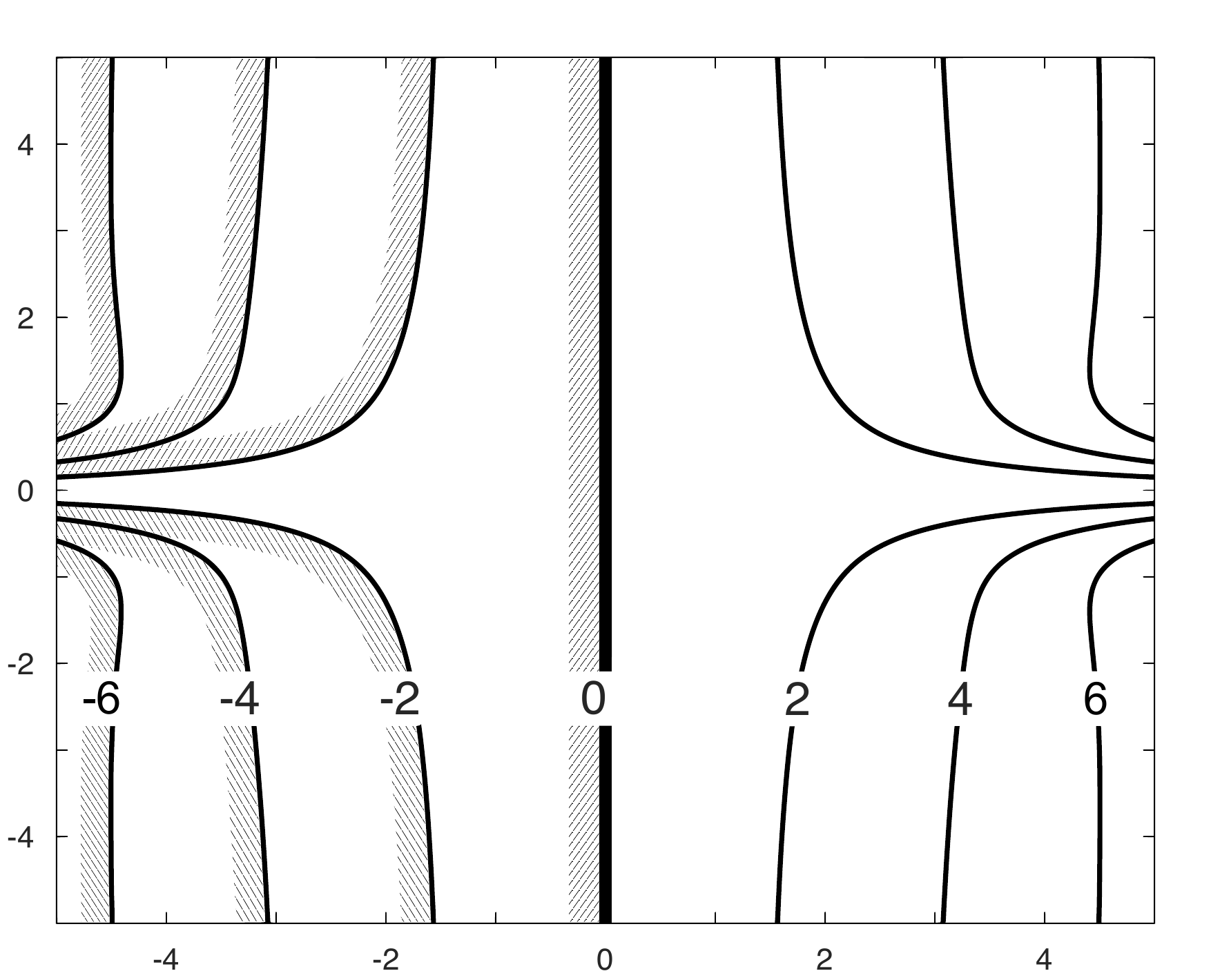}
  \caption{$\Im(\chi_{T1})$}
  \label{fig:bigsxwalk}
\end{subfigure}
\caption{Contour plots of the singulant branches of $\chi_{L1}$ and $\chi_{T1}$.
\label{fig:contours}}
\end{figure}

\noindent
In the case of Charpit's equations \eqref{lin-charpit}, solutions may be obtained by direct integration; this gives rays of the form
\begin{subequations}\label{rays-2d}
\begin{align}
x&=x_0+(2p_0+4p_0^3)\tau\label{xray-2d},\\
y&=y_0+2q_0\tau\label{yray-2d}, \\ 
\chi&= 2p_0^4\tau\label{chiray-2d}.
\end{align}
\end{subequations}
From \eqref{chiray-2d}, both solutions for $p_0$ yield 
\begin{subequations}
\begin{equation}\label{chi-analyt}
\chi = \frac{2h^4\tau}{s^4}.
\end{equation}
Moreover, rearrangement for $\tau$ in \eqref{xray-2d} produces
\begin{equation}\label{tau-analyt}
\tau =\mp \frac{s^3(s-x)}{2h(2h^2+s^2)},
\end{equation}
where the sign choice is written so as to produce branches consistent with \eqref{p0-q0-solns}.
Using the solutions \eqref{p0-q0-solns} for $p_0$ and $q_0$ along with \eqref{tau-analyt}
in \eqref{yray-2d} produces a quartic equation for $s$,
\begin{equation}\label{quartic}
(x^2 + y^2)s^4 + 4xh^2s^3 + (h^2x^2 + 4h^2y^2 + 4h^4)s^2 + 4h^4xs + (4y^2h^4 + 4h^6) = 0.
\end{equation}
\end{subequations}
This equation admits four solutions for $s$ which, along with the choice of sign in \eqref{tau-analyt} generate eight distinct branches of the singulant, $\chi$. We adopt the naming convention of \cite{lustri-2013}, labeling the four branches associated with \emph{longitudinal waves} $\chi_{L1},\dots,\chi_{L4}$, while those associated with \emph{transverse waves} are labeled $\chi_{T1},\dots,\chi_{T4}$. For brevity, we present free-surface contour plots of only two branches; one of longitudinal type and one of transverse type, which we label $\chi_{L1}$ and $\chi_{T1}$ respectively; see Fig.~\ref{fig:contours}. As may be readily seen from symmetries in the above terms, further branches are given by 
\begin{equation}
\begin{aligned}
\chi_{L2} &= \overline{\chi}_{L1}, & \chi_{T2} &= \overline{\chi}_{T1},\\
\chi_{L3} &= -\chi_{L1}, & \chi_{T3} &= -\chi_{T1},\\
\chi_{L4} &= -\overline{\chi}_{L1}, &  \chi_{T4} &= -\overline{\chi}_{T1},
\end{aligned}
\end{equation}
where the bar denotes complex conjugation.
\subsection{Physical validity of free-surface solutions}
  \noindent The radiation condition of \S\ref{sect:formulation} implies the presence of only the wave-free base solution \eqref{series-phi-eta} upstream of the point-source. The branches $\chi_{L1}, \chi_{L2}, \chi_{T3}$, and $\chi_{T4}$ satisfy the Dingle criterion [\emph{cf.} \eqref{eqn:dingle-2}] on the line $x=0$. However we note that
\begin{align}
\Re(\chi_{T3}),\, \Re(\chi_{T4})\to -\infty  \quad\text{when $y=0$, $x\to\infty$,}
\end{align}
and therefore if waves of the form $e^{-\chi/\ep}$ exist in this limit, they violate the condition of a bounded far-field. These branches are therefore precluded from being switched-on across $x=0$. Thus we conclude that the line $x=0$ is a Stokes line across which only the branches $\chi_{L1}$ and $\chi_{L2}$ are switched-on by the base solution \eqref{series-phi-eta}. We note the conjugacy of these branches implies real-waves on the surface. In addition to the switching of the longitudinal waves by the base series via \eqref{eqn:dingle-2}, in this problem we also have the occurrence of a \emph{higher-order Stokes line} where switching is caused by active exponential terms in the solution, this is then given by
\begin{equation}\label{eqn:ho-dingle}
\Im(\chi_{L1,2})=\Im(\chi_{T1,2}), \quad \text{and}\quad \Re(\chi_{T1,2})\geq \Re(\chi_{L1,2}).
\end{equation}
along the curve in Fig.~\ref{fig:regions}b. It is across this line that transverse branches $\chi_{T1}$ and $\chi_{T2}$ are switched-on by $\chi_{L1}$ and $\chi_{L2}$ respectively. We note that branches $\chi_{L3}, \chi_{L4}, \chi_{T3}$, and $\chi_{T4}$ remain dormant throughout. The solution schematic is presented in Fig.~\ref{fig:regions}b: the solution is waveless in region $\bm{A}$, composed of longitudinal waves (represented by branches $\chi_{L1}, \chi_{L2}$) in region $\bm{B}$, and both longitudinal and transverse waves (represented by $\chi_{L1}, \chi_{L2}, \chi_{T1},$ and $\chi_{T2}$) in regions $\bm{C}$.
\section{Asymptotics for far-field and shallow point-source}
\noindent
We now examine the behaviour of the singulant, $\chi$, in both the far-field limit and the limiting case in which the point-source approaches the free-surface. In both cases, we demonstrate an intimate connection between this problem and the classical Kelvin problem [see \cite{kelvin-1887}]. First, we show that in the far-field limit, $x\to \infty$, the higher-order Stokes line (shown in Fig.\ref{fig:regions}b) coincides with the Kelvin angle of $\arctan(1/{\sqrt{8}})\approx19.47^\circ$. Second, in the case of the point-source approaching the surface, $h\to 0$, and for any fixed $x>0$, we show that the higher-order Stokes line coincides with the Kelvin angle. This provides an extension to \cite{lustri-2013} who considered the case of $h=0$.
\subsection{Far-field limit}
\noindent
We now show that the higher-order Stokes line tends to the Kelvin angle in the far-field. Specifically, let us examine the far-field behaviour restricted to $y=\alpha x$ with $\alpha\in\mathbb{R}$. We seek the solution to the quartic equation, \eqref{quartic}, in asymptotic form 
\begin{equation}\label{eqn:s0-as-ser}
s = s_0 +\frac{s_1}{x}+\dots.
\end{equation}
At leading order, $\Oh(x^2)$, the quartic equation becomes
\begin{equation}\label{eqn:s0}
(1+\alpha^2)s_0^4+h^2(1+4\alpha^2)s_0^2+4h^4\alpha^2 =  0,
\end{equation}
and this admits four distinct solutions, given by
\begin{equation}\label{eqn:as-solns}
s_{01\pm}(\alpha)  = \pm\i h\frac{\sqrt{1+4\alpha^2+\sqrt{1-8\alpha^2}}}{\sqrt{2+2\alpha^2}}, \quad s_{02\pm}(\alpha)  = \pm\i h\frac{\sqrt{1+4\alpha^2-\sqrt{1-8\alpha^2}}}{\sqrt{2+2\alpha^2}}.
\end{equation}
It may be shown that solutions $s_{01\pm}(x,y)$ recover the far-field behaviour of longitudinal branches, while $s_{02\pm}(x,y)$ recover the behaviour of transverse branches. Thus when combined with the choice of sign for $\tau$ in \eqref{tau-analyt}, the far-field behaviours of all eight branches are captured by the series \eqref{eqn:s0-as-ser}. Let us denote the critical value as
 $$
\alpha^\star \equiv \frac{1}{\sqrt{8}},
$$
corresponding to the Kelvin angle. We see that at this value the branches degenerate to form only two distinct solutions for $s_0$,
\begin{equation*}
s_{01\pm}(\alpha^\star)  = \pm\i h\sqrt{\frac{2}{3}}, \quad s_{02\pm}(\alpha^\star) = \pm\i h\sqrt{\frac{2}{3}}.
\end{equation*}
Thus at the Kelvin angle we have that longitudinal and tranverse branches are of equal phase. Further, for $|\alpha|<\alpha^\star$, the real components of the singulant are,
\begin{equation*}
\Re\left[\chi(s_{01\pm})\right] =  \frac{2h(1+\alpha^2)}{5+8\alpha^2+\sqrt{1-8\alpha^2}}, \quad \Re\left[\chi(s_{02\pm})\right] =\frac{2h(1+\alpha^2)}{5+8\alpha^2-\sqrt{1-8\alpha^2}}.
\end{equation*}
 It may be readily seen that for $0<\alpha<\alpha^\star$, we have $\Re\left[\chi(s_{01\pm})\right]<\Re\left[\chi(s_{02\pm})\right]$, so it follows that the Dingle criterion is satisfied in the limit $|\alpha|\nearrow\alpha^\star$. Thus we conclude that in the far-field, $x\to\infty$, the higher-order Stokes line approaches the Kelvin angle, and moreover the higher-order Stokes line is confined to the inside of the Kelvin wedge (\emph{cf.} Fig.~\ref{fig:regions}).
\subsection{Shallow point-source limit}
\noindent
We now show that the higher order Stokes line coincides with the Kelvin wedge in the limiting case of a shallow point-source, $h\to0$. Following the same approach as above, we consider behaviour on the line $y=\alpha x$. We pose an asympotitic series for $s$ in ascending powers of the point-source depth parameter, $h$,  
\begin{equation}\label{s0-as-ser}
s = s_0 +hs_1+\dots.
\end{equation}
At leading order, $\Oh(1)$, the quartic equation \eqref{quartic} evaluates to
\begin{equation}
s_0^4x^2(1+\alpha) =  0.
\end{equation}
This produces the trivial leading-order solution, $s_0(x,y)=0$. The three subsequent orders [$\Oh(h), \Oh(h^2)$, and $\Oh(h^3)$] all vanish upon substitution of $s_0 =0$. The first nontrivial evaluation is at $\Oh(h^4)$, where we obtain the four solutions  
\begin{equation}
s_{41\pm}(\alpha)  = \pm\i\frac{\sqrt{1+4\alpha^2+\sqrt{1-8\alpha^2}}}{\sqrt{2+2\alpha^2}},\quad s_{42\pm}(\alpha)=\pm\i\frac{\sqrt{1+4\alpha^2-\sqrt{1-8\alpha^2}}}{\sqrt{2+2\alpha^2}}.
\end{equation}
By comparison of these terms with those in \eqref{eqn:as-solns}, we see that we may apply precisely the same arguments seen in the previous section. It follows that we may conclude that the higher order Stokes-line approaches the Kelvin angle as $h\to0$.
\section{Numerical computation of singulant branches}\label{sect:numerics}
\noindent

The key idea is that there is a link between $s$-space and $(x,y)$-space, this is a difficult to explore as the partitioning of $s$-space is highly dependent on the choice of parametrisation [\emph{cf.} \eqref{x-y-chi-ic}]. In order to handle this we design an algorithm which is explained as follows
 \begin{enumerate}[label=(\roman*),leftmargin=*, align = left, labelsep=\parindent, topsep=3pt, itemsep=2pt,itemindent=0pt ]
  \item Start with a parameter value $s=s_0$, and a particular choice of $(y_0\text{-sign}, p_0\text{-sign})$ [determining a branch of $y$ and $p$ via \eqref{x-y-chi-ic}]. These three choices determine a particular complex-ray.
  \item Generate data for $x,y,\dots,\chi$ on the finely meshed rectangular region in $\tau$-space by integrating Charpit's equations \eqref{lin-charpit} using any standard ODE solver (in our case \texttt{ode113} in \texttt{Matlab}). 
  \item Record the intersection of the zero-contours of $\Im(x)$ and $\Im(y)$, at $\tau = \tau^\star$.\footnote{For some special values of $s_0$, these contours will be parallel, but due to the particulars of our scheme, we shall not concern ourselves with this.}
  \item Calculate the values of $x^\star,y^\star,\dots,\chi^\star$ at the intersection point $\tau^\star$ via interpolation. This gives the value of $\chi$ at $(x^\star,y^\star)\in\mathbb{R}^2$.
\end{enumerate}
Upon repetition for many choices of $s=s_0$, the landscape of $\chi(x,y)$ emerges. We shall see shortly that different regions in $s$-space, along with different choices of $(y_0\text{-sign}, p_0\text{-sign})$ generate the eight different solution branches. For further insight into the inversion relationship $s\leftrightarrow(x,y)$, see App.~\ref{app:param} wherein we identify the natural partition of $s$-space according to this relationship (see Fig.~\ref{fig:mappings}). We note that while this analysis is helpful, it is not necessary for the numerical schemes of this section, and indeed for many problems such an analysis is not possible. With this in mind, we proceed as though the relationship is not known in advance. We now outline a continuation-like scheme by which we obtain solution data at sequential values of $s=s_0$. We refer to these collectively as a `\emph{walk}' and each such evaluation as a `\emph{step}' of the walk. With sufficient steps, each walk will comprehensively span one region in the $s$-space. For a visualisation of the relationship see Fig.~\ref{fig:mappings}. 
The numerical procedure is as follows: 
\begin{enumerate}[label=(\roman*),leftmargin=*, align = left, labelsep=\parindent, topsep=3pt, itemsep=2pt,itemindent=0pt ]
\item Select a value of $(y_0\text{-sign}, p_0\text{-sign})$ which shall remain fixed throughout, and pick an initial point in $s$-space, $s=s_0$. Choose some large value $L\in\mathbb{R}^+$ which will bound the walk by $\max(|x|,|y|) = L$.
\item Pick an initial step distance and direction and determine the values of $x^\star,y^\star, \chi^\star$ for this new point.
\item A change of sign in $x^\star,y^\star$, or a very large value of $\max(|x^\star|,|y^\star|)$ will inform us that have strayed outside of our desired region. In this case, interatively halve the step distance and repeat step (ii) until either the new point falls within the desired region, or we exceed some specified iteration limit.
\item If the iteration limit is exceeded we choose a new step direction. 
\item Proceed to walk in this new direction with original step distance.
\end{enumerate}
Using this procedure we are able to develop an understanding of the association between $s$- and $(x,y)$-spaces, and the results of this procedure are illustrated in Fig.~\ref{fig:mappings}. In particular, this visualises the process by which we may construct each of the singulant branches.
\section{Three-dimensional Stokes structure using complex-rays}\label{sect:3d}
\noindent
In this section, we outline the crucial ideas for how the main result of this work---the three-dimensional Stokes structure presented in Fig.~\ref{fig:regions}---may be generated. Later in \S\ref{sect:discussion} we shall discuss how these ideas may be extended to nonlinear problems. 
In this section, we extend the complex-ray method may of \S\ref{sect:complex} so that we may recover the singulant in three-dimensions, \emph{i.e.} away from the free-surface and within the physical fluid. Recall from \S\ref{sect:formulation} that within the fluid domain, $-\infty<z\leq0$, the singulant $\chi$ is governed by the Eikonal equation \eqref{chi-eik}. The motivation for solving on the free-surface was the lack of freely available initial data (see Sect.~\ref{sect:necessity}). However, we now demonstrate how suitable initial data for an analogous three-dimensional method is obtained.

 Now let us introduce
\begin{align}
\hat{p} &= \chi_{\hat{x}}, & \hat{q} &= \chi_{\hat{y}}, & \hat{r}  &= \chi_{\hat{z}},
\end{align}
 where, we introduce hats for the purpose of distinguishing the notation from free-surface data obtained by the method of \S\ref{sect:complex} [however, we note these represent the same spacial variables, in particular that $\hat{\chi}(\hat{x},\hat{y},0) = \chi(x,y)$]. Analogously to the previous section we permit our independent variables so be complex: $x,y,z\in\mathbb{C}$, and in accordance with complex-ray theory, the physical fluid domain is a real subspace. For brevity, we shall omit repetition of the details similar to those of \S\ref{sect:formulation}. Charpit's equations may be integrated directly, producing rays of the form
\begin{equation}
\begin{aligned}\label{3d-lin-rays}
\hat{x} &= 2\hat{p}_0t+\hat{x}_0, &  \hat{y} &= 2\hat{q}_0t+\hat{y}_0,& \hat{z} &= 2\hat{p}_0^2t+ \hat{z}_0, &\hat{\chi} &= \hat{\chi}_0,
\end{aligned}
\end{equation}
where we have used \eqref{eik-solve-cond} to eliminate $r_0$. We note that the parametric variable, $t$ is distinct from the parameter $\tau$ in \S\ref{sect:complex}. We choose the in initial curve at $t=0$ to correspond to $z=0$, the complexified free-surface. We write this in the form
\begin{equation}
\begin{aligned}\label{3d-ic}
\hat{x}_0 &= x(\tau,s), &  \hat{y}_0 &= y(\tau,s), &\hat{z}_0 &= 0, &\hat{\chi}_0 &= \chi(\tau,s),
\end{aligned}
\end{equation}
\begin{landscape}
%
\begin{minipage}{0.9\textwidth}
\begin{center}
\begin{preview}
    \begin{overpic}[width=0.9\textwidth]{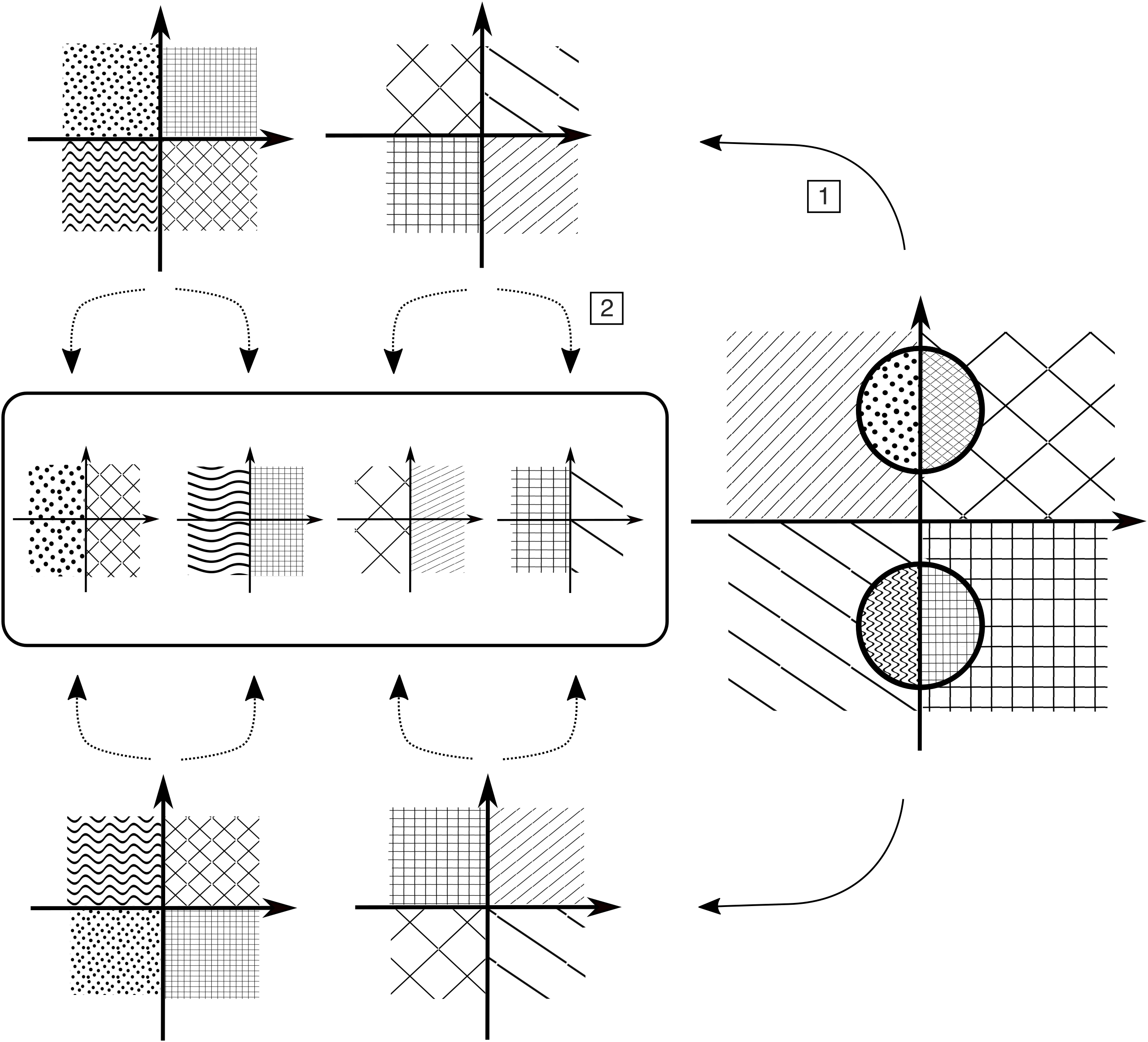}
    \put (75.5,23) { {$s$-plane}}
  \put (70,80) { {$(+,-)$}}
\put (70,8) { {$(-,+)$}}
        \put (5,53.5) {Reconstructed singulant branches}
         \put (5,37) { {$\chi_{L1}$}}
         \put (19,37) { {$\chi_{L2}$}}
         \put (33,37) { {$\chi_{T1}$}}
         \put (47,37) { {$\chi_{T2}$}}

    \end{overpic}
    \end{preview}
    \captionof{figure}{Recovery of the eight singulant branches by numerical complex-rays. Solid arrows denote mappings from the $s$-plane to $(x,y)\in\mathbb{R}^2$ with the labeled choice of $(y_0\text{-sign},p_0\text{-sign})$. Dotted arrows represent the combination of quadrants to recover each particular branch of the singulant $\chi$ (\emph{cf.} Fig.~\ref{fig:contours})\label{fig:mappings}}.
    
\end{center}
\end{minipage}
\hspace{0.5cm}
\begin{minipage}{0.5\textwidth}
\footnotesize For a given value of $s$ and a particular choice of $(y_0\text{-sign},p_0\text{-sign})$ [\emph{cf.} \eqref{x-y-chi-ic}, and \eqref{p0-q0-solns}],  there is a complex-ray which intersects real-space at some critical value of its parameter $\tau = \tau^\star(s)$. Thus substitution of $\tau^\star(s)$ into the ray equations \eqref{rays-2d} defines a map in which $s\mapsto (x,y,\chi)$. By the method outlined in \S~\ref{sect:numerics}, we may associate regions in $s$-space with their images in $(x,y)$-space under this mapping. In Fig.~\ref{fig:mappings}, this is visualised in step \fbox{$1$}. We find that in this manner, $s$-space is partitioned into eight regions, each corresponding to a quadrant in $(x,y)$-space. In this way, the correspondences between $s$-regions and $(x,y)$ quadrants are indicated with matching patterns in Fig.~\ref{fig:mappings}.
For a single choice of $(y_0\text{-sign},p_0\text{-sign})$, this process provides only partial information for a particular singulant branch. We must therefore repeat this process for more choices of $(y_0\text{-sign},p_0\text{-sign})$ to produce the remaining data and stitch together the results. This process is visualised in step \fbox{$2$}. For instance, data for the branch $\chi_{L1}$ in the second and fourth quadrants is obtained via rays corresponding to $(y_0\text{-sign},p_0\text{-sign})=(+,-)$ (step \fbox{$1$}), while we obtain data in the first and fourth quadrants via rays corresponding to $(y_0\text{-sign},p_0\text{-sign})=(-,+)$. The four quadrants are then combined appropriately (step \fbox{$2$}).
\end{minipage}%
\end{landscape}
where the right hand sides are given by rays \eqref{rays-2d}, and $\hat{p}_0$ and $\hat{q}_0$ are given by \eqref{p0-q0-solns}. We emphasise at this point the complex nature of the two-dimensional ray method of \S\ref{sect:complex}, in particular that the complex-rays allow us to obtain solution data for the entire complexified free-surface. Thus, initial data of the form \eqref{3d-ic} is available to us. In order to recover singulant in physical space, we now look for intersections of the three-dimensional complex-rays, \eqref{3d-lin-rays}, with real-space. This occurs when
\begin{equation}
\begin{aligned}\label{stokes-surf-cond}
\Im{\hat{x}}=\Im{\hat{y}}=\Im{\hat{z}}=0.
\end{aligned}
\end{equation}

For given initial conditions (\emph{i.e.} fixed $\tau$ and $s$), this represents an overdetermined system of three equations for two unknowns, $\Re(t)$ and $\Im(t)$. Thus, in general only rays originating from special points will intersect real space. The first two real-space conditions, $\Im(x)=0$ and $\Im(y)=0$, are satisfied by
\begin{equation}\label{eqn:t-soln}
t^\star(x_0,y_0,p_0,q_0) = -\frac{q_{01}x_{02}-p_{01}y_{02}}{2\left(p_{02}q_{01}-p_{01}q_{02}\right)}+\im \frac{q_{02}x_{02}-p_{02}y_{02}}{2\left(p_{02}q_{01}-p_{01}q_{02}\right)},
\end{equation} 
where we use notation of the form $x_0=x_{01}+\im x_{02}$.
To address the overdetermined nature of the system, we impose a requirement that the initial conditions be chosen in a particular manner. Substituting $t^\star(x_0,y_0,p_0,q_0)$ into the final real-space condition, $\Im(z)=0$, yields the requirement on the initial data,
\begin{equation}\label{eqn:z0-cond}
 \Im\left\{p_0^2\,t^\star\!\left[\hat{x}(\tau),\hat{y}(\tau),p_0,q_0\right]\right\}=0.
\end{equation}
This defines a contour in $\tau$-space. Each three-dimensional complex-ray originating from this contour will intersect real-$(x,y,z)$ space at $t = t^\star$. Thus each $s$ value produces a family of real-$(x,y,z)$ intersections. A visualisation of this process is presented in Fig.~\ref{fig:big}.

\section{Discussion}\label{sect:discussion}
\begin{landscape}
\begin{figure}
\begin{overpic}[width=1.4\textwidth]{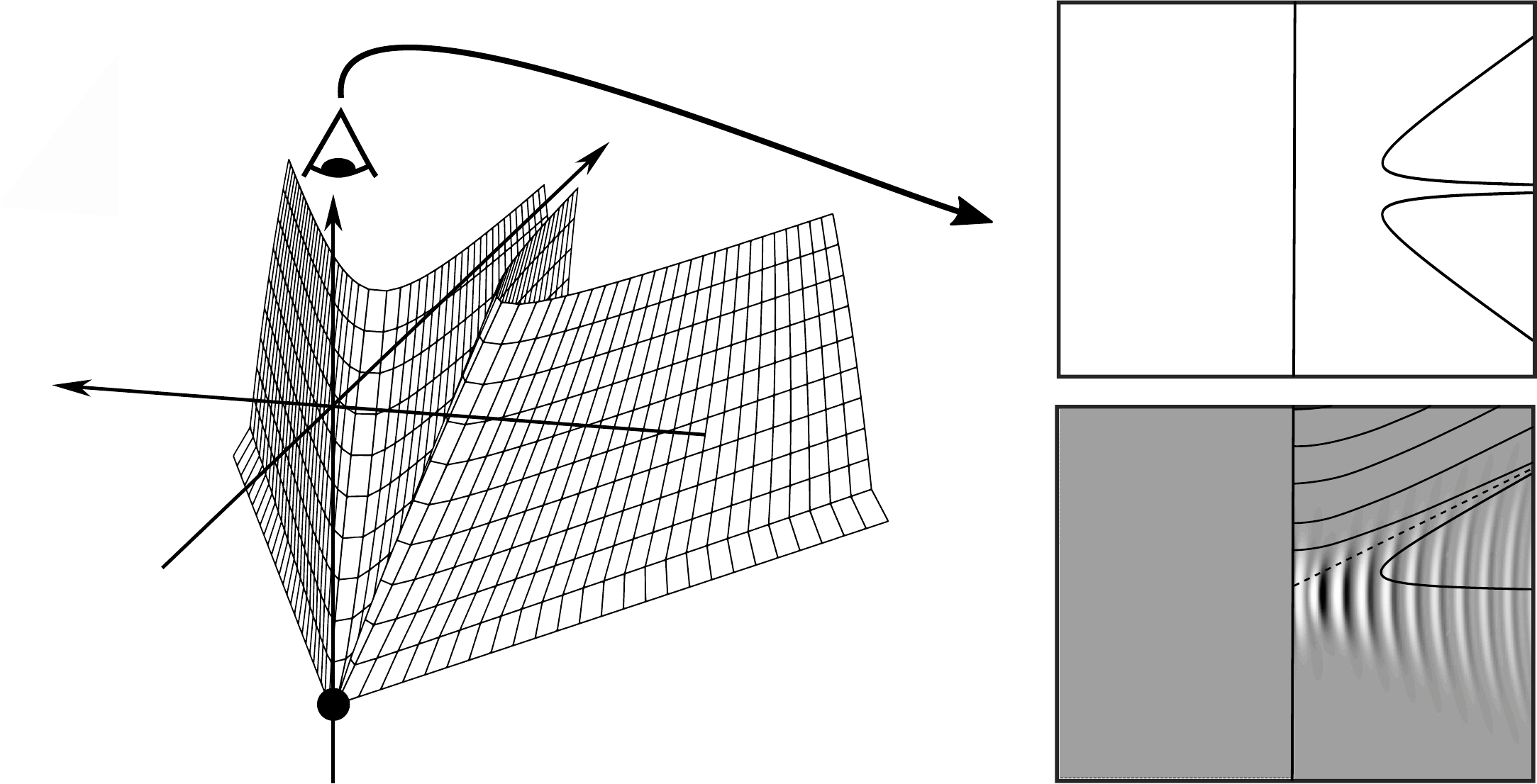}
    \put (78,38) {{$\bm{A}$}}
    \put (86,38) {{$\bm{B}$}}
    \put (95,35) {{$\bm{C}$}}
    \put (95,41) {{$\bm{C}$}}
    \put (95,41) {{$\bm{C}$}}
    \put (91,40.5) {{$\otimes$}}
    \put (88,39.5) {{$\oplus$}}
    \put (40,41) {{$x$}}
    \put (5,27) {{$y$}}
    \put (23,37) {{$z$}}
    \put (87,18.2) {\rotatebox{25}{$\Re(\chi_L)=\text{const.}$}
    }
  
       \end{overpic}
        \caption{ (a): Longitudinal branches are switched-on across the Stokes surface defined by $x=0$ separating regions $\bm{A}$ and $\bm{B}$, and switch-on transverse branches across the higher-order Stokes surface (meshed), separating regions $\bm{B}$ and $\bm{C}$. (b): The solution is composed of regions in which the solution has no waves ($\bm{A}$), longitudinal waves ($\bm{B}$), and both longitudinal and transverse waves ($\bm{C}$). 
      (c): Solution schematic imposed over the numerical free-surface.\label{fig:regions}}
\end{figure}
\end{landscape}

\noindent This paper was driven by two principal motivations. First was to extend to the work of \cite{lustri-2013} who had studied linearised flow past a submerged source in the low-Froude limit and uncovered the Stokes structure---that is, the family of Stokes lines across which exponentially small waves \emph{switch-on}---on the free-surface. Additionally, we wished to develop a method which would be applicable to a wider class of problems than linearised gravity flow. The work of \cite{lustri-2013} relied on a complex-ray approach in which the linear PDEs governing the problem were solved analytically, with the respective solution branches revealing the nature of the Stokes structure. These results naturally provoke the question of the existence of \emph{Stokes surfaces}, the three-dimensional analogue of Stokes lines, within the fluid. The nature of such structures was not addressed by \cite{lustri-2013}, and indeed the two-variable complex-ray approach utilised by the authors for the free-surface problem is not sufficient to reveal Stokes phenomena within the fluid. Our aim, therefore, was to develop a method which both replicates the results of \cite{lustri-2013} on the fluid free-surface and reveals the three-dimensional Stokes structure. In this paper, we have reproduced these findings numerically, and have developed a numerical method which confirms what physical intuition would suggest; that there exists a similarly sophisticated Stokes structure within the fluid itself. We believe Fig.~\ref{fig:regions} is the first visualisation of such three-dimensional structures. 
\par
\emph{Is it necessary to resort to complex-ray theory?} In essence, the singulant is governed by a first order boundary-value problem (with boundary constraints given by \eqref{eqn:ic-intro} along with \eqref{eqn:bc-intro} on $z=0$). Crucially these curves are themselves partially intersecting. In order to ensure both conditions are satisfied, we intitialise our method at the intersection
\begin{landscape}
\vspace*{1cm}
\begin{center}
\begin{overpic}[width=0.9\textwidth]{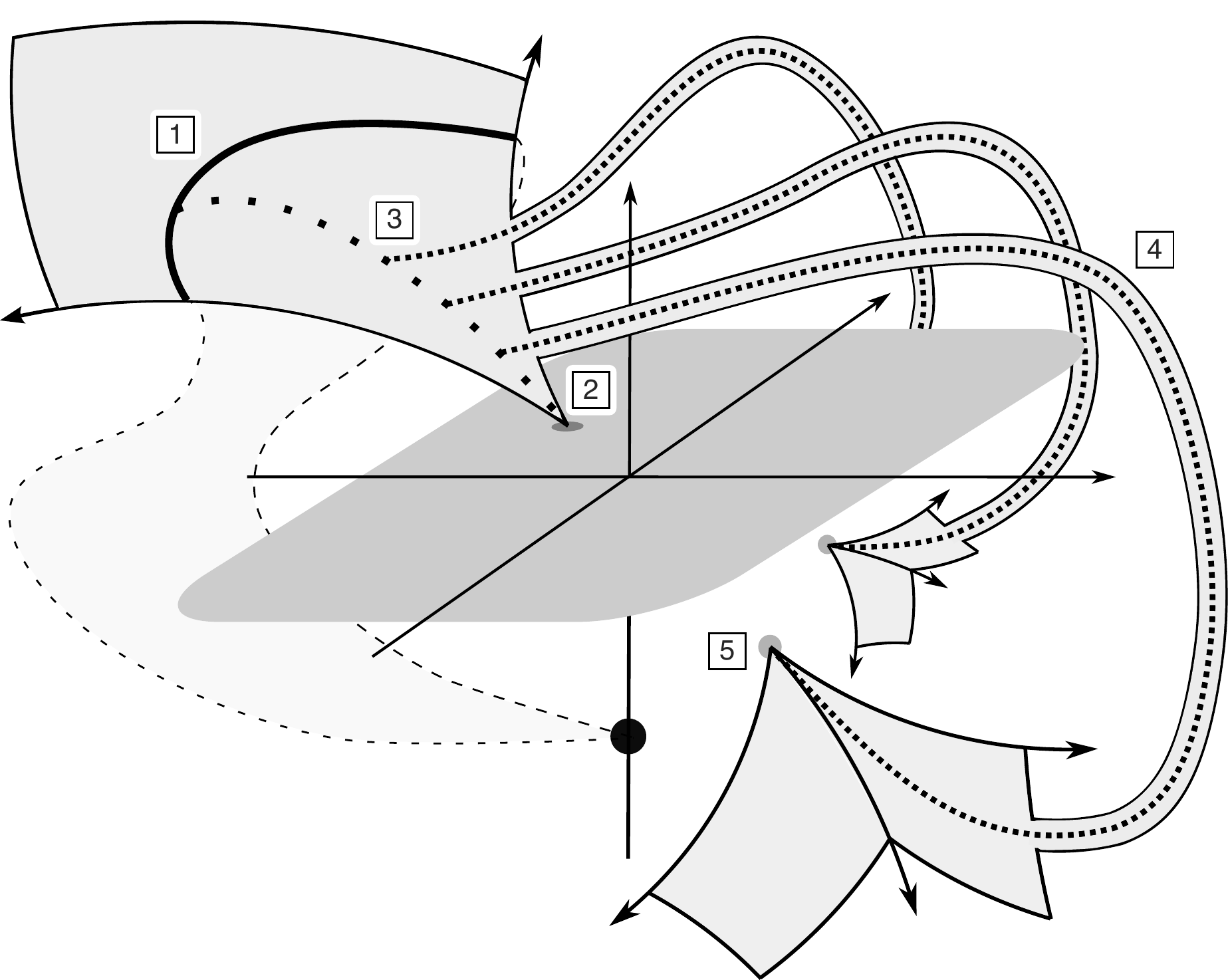}
    \put (52.5,64) {\scriptsize $\Re(z)$}
    \put (63,54) {\scriptsize $\Re(x)$}
    \put (89,42) {\scriptsize$\Re(y)$}
  \put (45,74) {\scriptsize$\Im(x)$}
  \put (3,52) {\scriptsize$\Im(y)$}
       
       \put (86,20) {\scriptsize$\Im(y)$}
       \put (75,3) {\scriptsize$\Im(z)$}
       \put (45,7) {\scriptsize$\Im(x)$}
       \put (-14,85) {\parbox{7cm}{\footnotesize \framebox(2.5,2.5){1}\,: data is known on the initial curve, $x^2+y^2+h^2=0$, indicated by bold line. A two dimensional complex-ray propagates through $(x,y)\in\mathbb{C}^2$ space (\emph{cf.} \S\ref{sect:complex}).}}
       \put (-38,44) {\parbox{5cm}{\footnotesize \framebox(2.5,2.5){2}\,: The value of $\chi(x,y,0)$ is recovered at the $\mathbb{R}^2$-space intersection.}}
      \put (-22,9) {\parbox{7cm}{\footnotesize \framebox(2.5,2.5){3}\,: In order to recover the singulant value in $(x,y,z)\in\mathbb{R}^3$ space, initial data is harvested from the dotted curve given in \eqref{eqn:z0-cond}. }}
      \put (75,76) {\parbox{7cm}{\footnotesize \framebox(2.5,2.5){4}\,: Complex-rays propagate through $(x,y,z)\in\mathbb{C}^3$ space (\emph{cf.} \S\ref{sect:3d}). }}
      \put (98,6) {\parbox{4cm}{\footnotesize \framebox(2.5,2.5){5}\,: From each complex-ray, a singulant value at a point in the physical fluid is recovered at the $\mathbb{R}^3$-space intersection.}}
       \end{overpic}
       \captionof{figure}{A visualisation of the complex-ray procedures presented in \S\ref{sect:complex} and \S\ref{sect:3d}. The physical free-surface is indicated by the horizontal plane with grey shading. When the solution at a given point is analytically continued we visualise this as an embedded plane ($\mathbb{R}^2$) or volume ($\mathbb{R}^3$) unfurling from the relevant point. \label{fig:big}}.
      \end{center}
\end{landscape}

 of these two boundaries in complex-space; meaning that any ray scheme will necessarily require complex-rays. We note also the importance in understanding the relationship between solutions in complex-space and those in real-space. For two-dimensional problems, the utility of complex-rays is well documented [see \emph{e.g.} \cite{complex-rays-book}]. In such contexts, a single complexified independent variable may be readily associated with real space by the natural mapping $\mathbb{C}\to \mathbb{R}^2$ [see \emph{e.g.} \cite{crew-trinh-2016}]. In higher dimensions, the ambiguous relationship between $\mathbb{C}^2$ and $\mathbb{R}^3$ makes the complexification of two or more variables considerably more complicated. 
\par
 \emph{What other problems can be studied using this complex-ray approach?}
  \noindent Our method may in principle be applied to any generic problem governed by a first order PDE with sufficient known data along a curve. In particular, the method is well suited to problems requiring the solution of a singulant function as we know it must vanish on the curve along which the velocity potential is singular. Thus, a natural extension to this work is to the problem of linearised gravity-capillary flow over a submerged source, where the governing equation in place of \eqref{chi-prob} is
  $$
  \beta^2\chi_x^4+(\chi_x^2+\chi_y^2)[\beta\tau(\chi_x^2+\chi_y^2)-1]^2=0,
  $$ 
  where $\beta$ and $\tau$ are associated with the Froude and Weber numbers respectively [see \cite{lustri_pethiyagoda_chapman_2019}]. \par
\begin{figure}
  \centering
  \includegraphics[width=.6\linewidth]{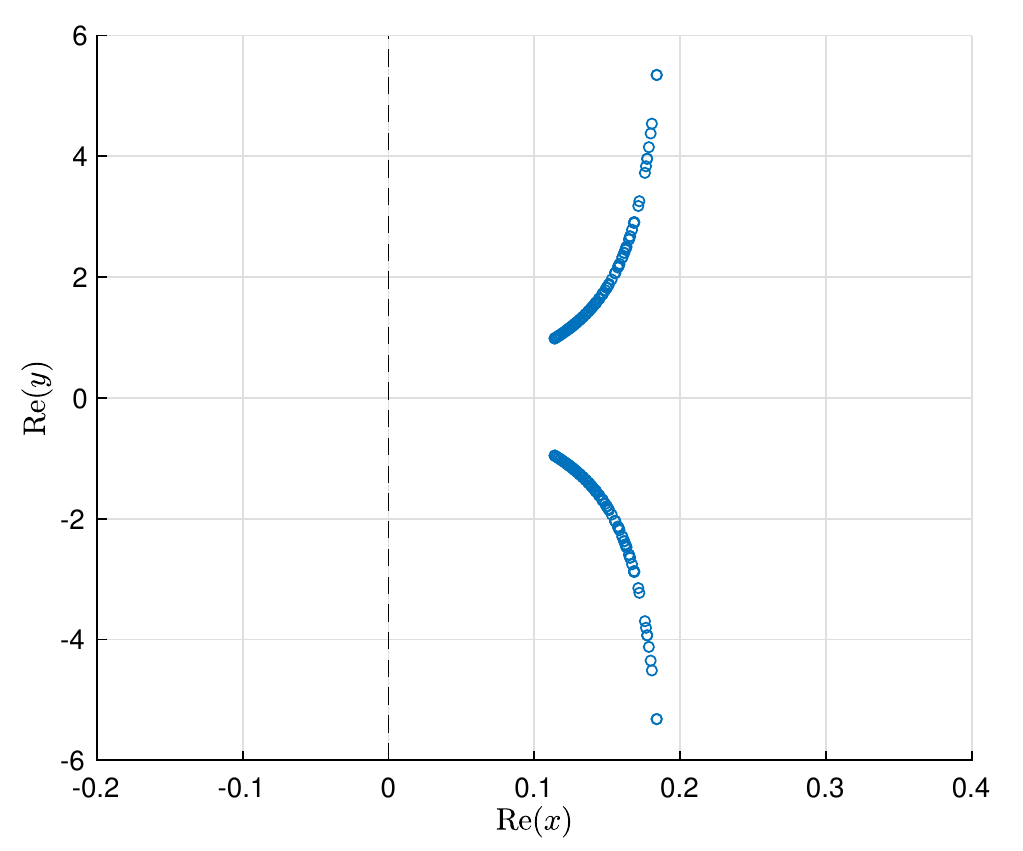}
\caption{The intersection of the Stokes surface with the free-surface for the problem of nonlinear flow over a point-source of $\Oh(1)$ strength.\label{fig:nonlin}}
\end{figure}

 Despite the method making no assumption of linearity in the underlying equations, the recovery of the singulant for the nonlinear problem is nontrivially more complicated, with the corresponding analysis generating the free-surface solvability condition
\begin{equation*}\label{3d-solve-cond}
\chi_z=(\chi_x{\phi_0}_x+\chi_y{\phi_0}_y)^2
\end{equation*}
in place of  \eqref{3d-solve-cond}. This produces the governing equation for $\chi$,
\begin{equation*}\label{3d-phi-prob}
\chi_x^2+\chi_y^2+(\chi_x{\phi_0}_x+\chi_y{\phi_0}_y)^4=0,
\end{equation*}
leading to a considerably more complicated problem due to the governing equations containing derivatives of a singular $\phi_0$. This problem is not amenable to analytical solution, however, the numerical complex-ray methods developed in this work may in principle be applied to this more general problem. In particular, the three-dimensional portion of the complex-ray scheme of \S\ref{sect:3d} relies only on the Eikonal equation \eqref{chi-eik}, a linear PDE.
  Thus, this crucial part of the method remains simple even when studying the problem without linearisation assumptions.
  In all cases, the logistical difficulty of the stitching procedure outlined in \S\ref{sect:numerics} depends upon the simplicity of correspondence between physical space and parameter space. In linear problems, this may be established analytically. However, for nonlinear problems, such a-priori information is not available, thus an application of the numerical method of \S\ref{sect:numerics} to such problems will likely necessitate a fine meshing of the parameter space.
 In Fig.~\ref{fig:nonlin} we present one preliminary numerical result from the nonlinear submerged point-source problem, namely the Stokes line on the free-surface, \emph{i.e.} the intersection of the Stokes surface with the free surface [see \cite{fitzgerald-2018}]. 

\emph{May the Fourier analysis method be applied to nonlinear problems?}
The efficacy of the Fourier approach of \S\ref{app:fourier} relies on the linearity of the underlying governing equations. For the nonlinear problem described above, reformulation of corresponding governing equations in terms of Fourier variables requires the use of convolutions, greatly exacerbating the process of obtaining Fourier solutions. \par
\emph{May the methods developed in this paper be used to model flow around three-dimensional bluff bodies?} It is known [see \cite{liu-tao-2001}] that more general bodies may be modeled using a composition of point-sources. A firm understanding of the nonlinear point-source problem, underpinned by the methods of this work may extended to a theory for more general nonlinear obstructions. 
\appendix
\section{Analysis of the parameter space}\label{app:param}
\noindent
\noindent Recalling that Charpit's equations \eqref{lin-charpit} imply that $p\equiv p_0$ and $q\equiv q_0$, let us define
\begin{align}
X(s)&\coloneqq \dd{x}{\tau}= 2p_0(s)+4p_0^3(s), & &\text{and} & Y(s)&\coloneqq \dd{y}{\tau} = 2q_0(s). 
\end{align}
In this linear problem, Charpit's equations \eqref{lin-charpit} may be solved explicitly, giving rays
\begin{align}\label{lin-rays}
x(\tau;s)&=x_0(s)+X(s)\tau, & &\text{and} & y(\tau;s)&=y_0(s)+Y(s)\tau,
\end{align}
We see that these rays intersect real-$(x,y)$ space when 
$$
\tau = \tau^\star(s) = \frac{1}{D}\left(X_1\Im{y_0}-Y_1\Im{x_0}+\i(Y_2\Im{x_0}-X_2\Im{y_0})\right),
$$
where $X = X_1+\i X_2$, $Y = Y_1+\i Y_2$, and $D(s)\coloneqq X_2Y_1-X_1Y_2$. Thus the intersection points of the complex-ray in real-$(x,y)$ space are $\left(x^\star(s), y^\star(s)\right)=\left(x(\tau^\star;s), y(\tau^\star;s)\right)$. At this point, we note there is a natural partitioning of $s$-space by the curves along which $|x^\star(s)|+|y^\star(s)| \to \infty$. This partition produces eight sub-regions of $s$-space, labeled as in the centroid of Fig.~\ref{fig:mappings}. As it transpires, these regions recover distinct branches of the singulant, $\chi$, when we construct their images in $(x,y)$-space in the manner illustrated in Fig.~\ref{fig:mappings}.
\subsection{Observations on the inversion relationship $s\leftrightarrow(x,y)$}
\begin{enumerate}[label=(\roman*),leftmargin=*, align = left, labelsep=\parindent, topsep=3pt, itemsep=2pt,itemindent=0pt ]
  \item  The $s$-plane is partitioned into eight distinct regions of interest seperated by the zero contours of $1/{x^\star(s)}$ and $1/{y^\star(s)}$. we label these regions $I_1,\dots,I_4, O_1,\dots,O_4$ (as given in Fig.~\ref{fig:mappings}). 
  \item Regions $I_1,\dots,I_4$ produce branches of the singulant of a longitudinal type ($\chi_L$), while regions $O_1,\dots,O_4$ produce those of a transverse type ($\chi_L$ and $\chi_T$ respectively, Fig.~\ref{fig:contours}).
  \item For each branch of the singulant, $\chi$, the quadrants in $(x,y)$-space are composed as follows
  \begin{enumerate}
    \item Vertically adjacent quadrants correspond to the same region in $s$-space via complex-rays with opposing choices of ($y_0$-sign, $p_0$-sign).
  \item Diagonally opposite quadrants correspond to vertically adjacent region sin $s$-space via complex-rays with matching choices of ($y_0$-sign, $p_0$-sign).
    \end{enumerate}
\end{enumerate}
\section{Examination by Fourier analysis and steepest descent paths}\label{app:fourier}
\noindent We now provide a brief overview of an alternative derivation of the singulant using Fourier analysis, and show how the Stokes phenomenon may be realised by studying the steepest descent approximations of the respective Fourier integrals. In the context of steepest descents, the Stokes phenomenon corresponds to a sudden change in saddle contributions to the steepest descent path; it is in this context that the Dingle criterion \eqref{eqn:dingle-general} can be most readily intuited. Our Fourier analysis follows the methods found in \cite{noblesse-1981}, \cite{hermans-2011}, and others. We study higher-order terms, collectively $\psi$, in the velocity potential expansion seen in \S\ref{sect:formulation},
\begin{equation}
\phi=-\frac{1}{4\pi\sqrt{x^2+y^2+(z-h)^2}}-\frac{1}{4\pi\sqrt{x^2+y^2+(z+h)^2}}+\psi.
\end{equation}
Following \cite{lustri-2013} we re-formulate the problem in terms of Fourier variables $k$ and $l$, leading to an integral form for the velocity potential,
\begin{align}
\psi(\bm{x}) &= \frac{\ep}{4\pi^2}\int^{\infty}_{-\infty}\int^{\infty}_{-\infty}\frac{k^2e^{\rho(z-h)}}{\rho(\rho-\ep k^2)}e^{\im kx+\im ly}\de k\de l,\\
\intertext{where $\rho=\sqrt{k^2+l^2}$. We note the presence of integrand singularities on the real $k$-axis; the difficulty in handling inverse Fourier transforms with singular integrands in a well defined manner is addressed in \cite{noblesse-1981, eggers-1992, hermans-2011}, and others. In particular, \cite{noblesse-1981} outlines how a variety of equivalent analyses which may be performed using different but equivalant forms of the above integral expression, and compares their relative merits. One such representation is obtained by transforming to polar coordinates and setting $\rho = u/{\cos^2(\varphi)}$, producing the expression} 
\psi(\bm{x}) &= \frac{1}{4\pi^2}\int^{2\pi}_{0}\int^{\infty}_{0}\frac{\ep u}{(1-\ep u)\cos^2(\varphi)}e^{K(\varphi; \bm{x})u}\de u\de \varphi,
\end{align}
with the exponent given by
$$
K(\varphi;\bm{x}) = \sec^2(\varphi)\left[\im r \cos(\varphi-\theta)+(z-h)\right].
$$

 \noindent In this form, the singular behaviour of the integrand is consolidated into a single simple pole at $u = \ep^{-1}$ on the real $u$-axis. Following \cite{noblesse-1981}, \cite{hermans-2011}, and others, the radiation condition may be satisfied by appropriate consideration of the integral contribution from this pole. If $\cos(\varphi-\theta)>0$, to ensure convergence we close the inner integral in the first quadrant, indenting so as to enclose the pole if $x>0$, and exclude the pole if $x<0$. Conversely if $\cos(\varphi-\theta)<0$ we close the integral in the fourth quadrant and apply the opposite indentation procedure [\emph{cf.} \cite{hermans-2011} p.37]. Accordingly, there is a natural decomposition of the velocity potential into what \cite{noblesse-1981} refers to as \emph{near-field disturbance}, and \emph{wave disturbance},
$$
\psi(\bm{x}) \sim N(\bm{x}) + W(\bm{x}).
$$
The \emph{near-field disturbance} is given by
\begin{align}
N(\bm{x}) &=  \frac{1}{4\pi}\left(\int^{2\pi}_{0}\dashint^{\infty}_{0}-\int_{\mathcal{C}_1}\int^{+\infty\im}_{0}+\int_{\mathcal{C}_2}\int^{0}_{-\infty\im}\right)\sec^2(\varphi)\frac{\ep u}{(1-\ep u)}e^{Ku}\de u \de \varphi,\\
\intertext{while so-called \emph{wave disturbance} oscillatory contributions are contained in the term} 
W(\bm{x}) &= \frac{\im H(x)}{2\ep}\left(\int_{\mathcal{C}_1}-\int_{\mathcal{C}_2}\right)\sec^2(\varphi)e^{K/\ep} \de\varphi.\label{eqn:W1}
\end{align}
The contours of integration are given by
$$
\mathcal{C}_1 = \left\{\varphi\mid 0\leq \varphi < 2\pi, \cos(\varphi-\theta)>0\right\},
$$
$$
\mathcal{C}_2 = \left\{\varphi\mid 0\leq \varphi < 2\pi, \cos(\varphi-\theta)<0\right\},
$$
while $H(x)$ denotes the Heaviside unit-step function.
\subsection{Approximating the oscillatory integral by steepest descents}\label{sect:steps}
 \noindent In the asymptotic limit $\ep\to 0$ the exponentially small terms seen in \S\ref{sect:formulation} may be obtained via the method of steepest descents. For a comprehensive review of the steepest descents method, see \emph{e.g.} \cite{bleistein-1986}. In essence, it is an asymptotic technique which enables the approximation of certain integrals by the evaluating only their contributions at certain critical points; in particular branch singularities and \emph{so-called} saddle points. As applied to the oscillatory term, $W(\bm{x})$, we may expect saddle point contributions from stationary values of the integrand exponent of \eqref{eqn:W1}, \emph{i.e.} at the zero locations, $\varphi_s$, of
\begin{equation}
K^\prime(\varphi) = \frac{\im r \left[\sin(2\varphi-\theta)+3\sin(\theta)\right]+4\sin(\varphi)(z-h)}{2\cos^3(\varphi)}.
\end{equation}
Each critical point has an associated steepest descent contour, following $\Im(K)=\text{const.}$ and such that $\Re(K)$ is maximised at $\varphi_s$. The goal of the method is to replace the original integration contour with a path composed of one or more contours of steepest descent according to Cauchy's theorem. Thus, a saddle point contributes to the final integral approximation if and only if it is both possible and necessary for the steepest descent path to contain the contour of that particular saddle; we note that integration end points always contribute. For each such critical point, it may be shown [see \cite{bleistein-1986}] that the integral contributions are proportional to
\begin{equation}\label{eqn:saddleterm}
A(x,y,z)\exp\left[\frac{K\left(\varphi_s(x,y,z)\right)}{\ep}\right].
\end{equation}
 By comparing this with the exponentially small terms \eqref{rem-exp}, we note the natural association between the saddle point exponents, $K(\varphi_s(x,y,z))$, and the singulant, $\chi(x,y,z)$. Although the exact locations of the saddles are not explicitly soluble, we may leverage conjugate symmetries in the exponent, $K(\varphi(x,y,z))$, to show that a saddle located at $\varphi=\varphi_s$ implies the location of a further saddle point at $\overline{\varphi}_s+\pi$. To see this, we note the relation
$$
K^\prime(\overline{\varphi}+\pi) = \frac{-\im r \left[\sin(2\overline{\varphi}-\theta)+3\sin(\theta)\right]+4\sin(\overline{\varphi})(z-h)}{2\cos^3(\overline{\varphi})}=\overline{K^\prime(\varphi)}.
$$
Furthermore, $K$ satisfies the relationship $K(\overline{\varphi}+\pi)=\overline{K(\varphi)}$, and so the branches corresponding to the saddles $\varphi_s(x,y,z)$ and $\overline{\varphi}_s(x,y,z)+\pi$ are conjugate pairs. Indeed we see the same conjugacy relations between saddle contributions as those between singulant branches (\emph{cf.} \S\ref{sect:algebraic}).
 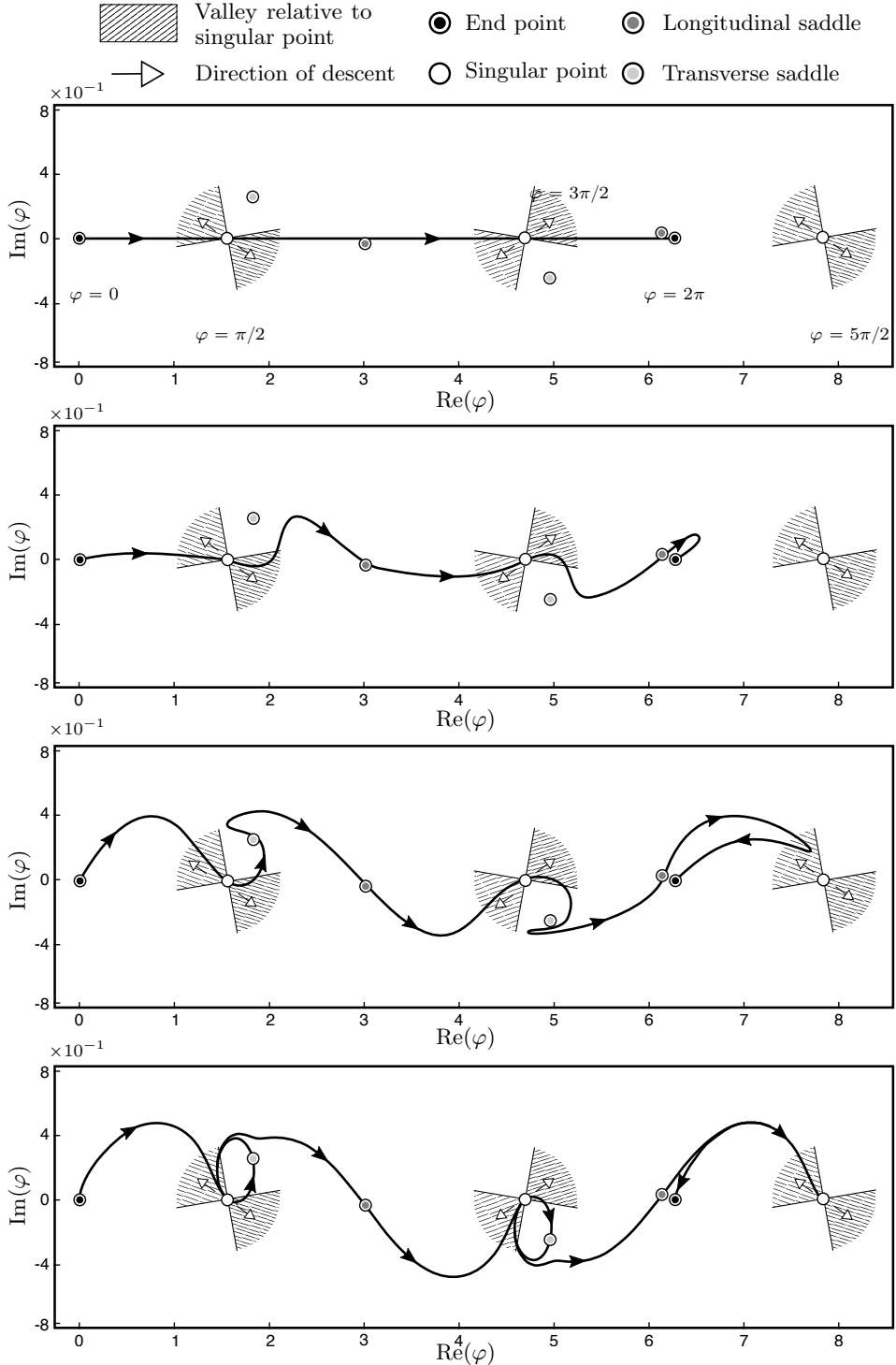
\begin{figure}
\centering
   \begin{overpic}[width=0.9\textwidth]{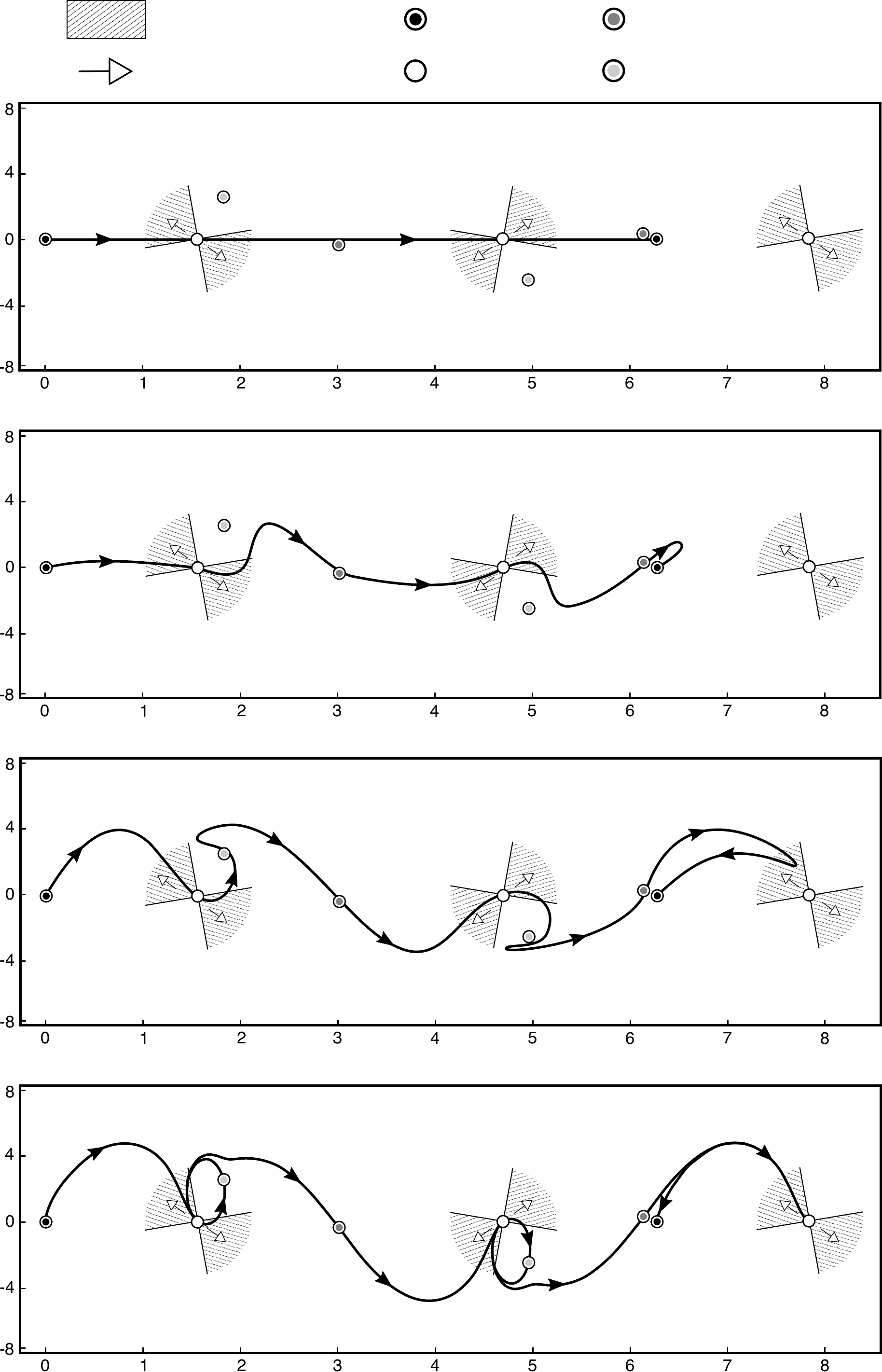}
   \put (1,93) {\scriptsize{$\times 10^{-1}$}}
   \put (1,69.1) {\scriptsize{$\times 10^{-1}$}}
   \put (1,45.6) {\scriptsize{$\times 10^{-1}$}}
   \put (1,21.6) {\scriptsize{$\times 10^{-1}$}}
   \put (2,78) {\scriptsize\colorbox{white}{$\varphi=0$}}
   \put (45,78) {\scriptsize\colorbox{white}{$\varphi=2\pi$}}
   \put (12,75) {\scriptsize{$\varphi=\pi/2$}}
   \put (37,85.5) {\scriptsize{$\varphi=3\pi/2$}}
   \put (58,75) {\scriptsize{$\varphi=5\pi/2$}}
   \put (12,98) {\parbox{2.5cm}{\footnotesize Valley relative to singular point}}
   \put (12,94.5) {\parbox{5cm}{\footnotesize Direction of descent}}
   \put (32.2,98.1) {\parbox{5cm}{\footnotesize End point}}
   \put (32.2,94.5) {\parbox{5cm}{\footnotesize Singular point}}
   \put (47,98.1) {\parbox{5cm}{\footnotesize Longitudinal saddle}}
   \put (47,94.5) {\parbox{5cm}{\footnotesize Transverse saddle}}

   \put (30,70) {\parbox{5cm}{\footnotesize$\Re(\varphi)$}}
   \put (30,46.2) {\parbox{5cm}{\footnotesize$\Re(\varphi)$}}
   \put (30,22.5) {\parbox{5cm}{\footnotesize$\Re(\varphi)$}}
   \put (30,-1) {\parbox{5cm}{\footnotesize$\Re(\varphi)$}}

   \put (-2,81) {\rotatebox{90}{\footnotesize$\Im(\varphi)$}}
   \put (-2,57) {\rotatebox{90}{\footnotesize$\Im(\varphi)$}}
   \put (-2,32.5) {\rotatebox{90}{\footnotesize$\Im(\varphi)$}}
   \put (-2,9) {\rotatebox{90}{\footnotesize$\Im(\varphi)$}}
\end{overpic}
\caption{A continuous deformation of the integration contour $[0,2\pi]$ into the path of steepest descent for a point in region $\bm{C}$, $(x,y,z)\approx(8.29,1.11,0)$ [\emph{cf.} Fig~\ref{fig:regions}].\label{fig:deformation}}
\end{figure}

\subsection{The higher-order Stokes phenomenon}
\begin{figure}
\centering
\begin{subfigure}{.32\textwidth}
  \centering
  \includegraphics[width=.8\linewidth]{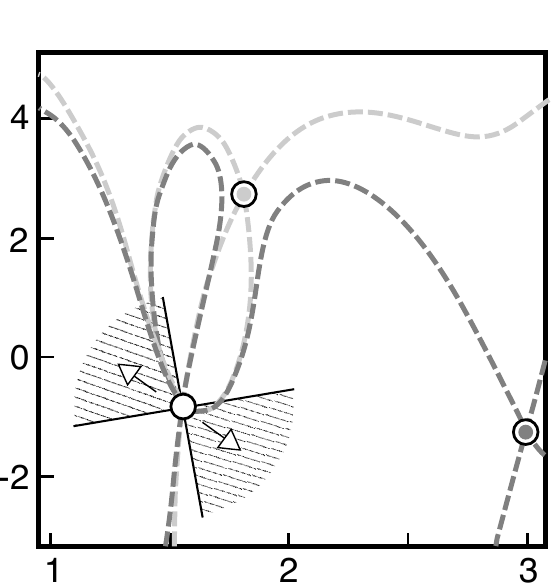}
  \caption{$(x,y)\approx(6.71,0.89)\in\bm{B}$.}
\end{subfigure}%
\begin{subfigure}{.32\textwidth}
  \centering
  \includegraphics[width=.8\linewidth]{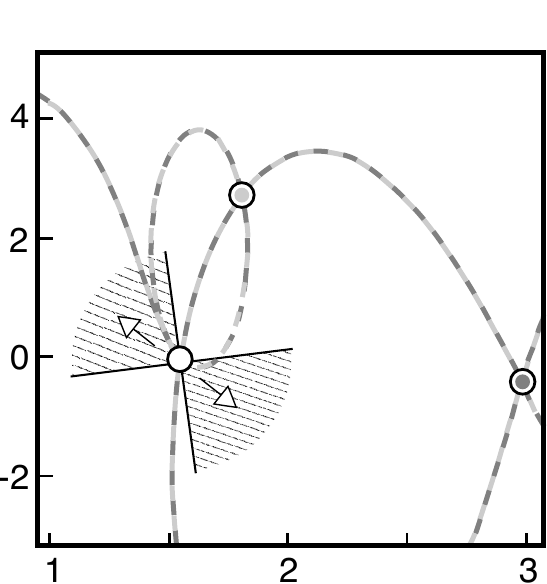}
  \caption{$(x,y)\approx(7.48,0.99)$}
\end{subfigure}%
\begin{subfigure}{.32\textwidth}
  \centering
  \includegraphics[width=.8\linewidth]{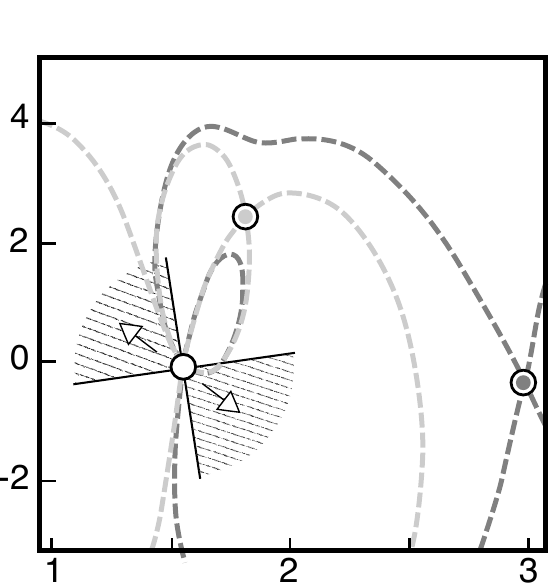}
  \caption{$(x,y)\approx(8.29,1.11)\in\bm{C}$.}
\end{subfigure}%
\caption{Equal phase contours (paths of steepest descent/ascent) for locations slightly left of, on, and slightly right of the higher-order Stokes line. \label{fig:transition}}
\end{figure}

\noindent As emphasised in \cite{bleistein-1986}, a crucial step in the method of steepest descents is a justification for the replacement of the integration contour with the final steepest descents path. In practice, this is satisfied if we can perform a continuous deformation between the paths. For the integral evaluated at $(x,y,z)\approx(8.29,1.11,0)$, indicated by $\otimes$ in Fig.~\ref{fig:regions}, a sketch of such a deformation is presented in Fig.~\ref{fig:deformation}. We note that this point is inside region $\bm{C}$, thus we expect saddle point contributions of both longitudinal and transverse type. Indeed, the final steepest descent path passes through both of these saddle types.
  By comparing the steepest descent contours in this case with those in which the integration is performed at a location in region $\bm{B}$, it may be verified that it is unnecessary to proceed through transverse saddle points and thus waves of this type remain inactive here. In Fig~\ref{fig:transition}, we compare the equal-phase contours, $\Im(K(\varphi))=\Im(K(\varphi_s))$, as we traverse the higher-order Stokes line transition from region $\bm{B}$ into region $\bm{C}$, and note the transition of these contours as the higher-order Stokes line is crossed.

\subsubsection{Local analysis of branch singularities}\label{app:singularity}
\noindent We note the presence of critical points of the integrand when $\cos(\psi)=0$ \emph{i.e.}
\begin{equation}\label{eqn:sing}
\varphi_{0n} = (n+1/2)\pi.
\end{equation} 
 To study the local behaviour of the exponent function, $K$, we introduce 
 \begin{align}
Me^{\im m} &= \im r \cos(\pi/2-\theta)+z-h,\\
\rho e^{\im\sigma} &= \varphi - \pi/2.
\end{align}
In the vicinity of these singular points, we have that 
\begin{align}
K&\sim\frac{M}{\rho^2}e^{\im(m-2\sigma)}\\
&= \frac{M}{\rho^2}\sin(m-2\sigma) +\im\frac{M}{\rho^2}\cos(m-2\sigma),
\end{align}
From this we see that equal phase contours $\Im(K) = C$ [\emph{i.e.} paths of steepest descent/ascent in $\Re(K)$] are given by 
\begin{equation}
\rho = \sqrt{\frac{M}{C}\sin(m-2\sigma)}.
\end{equation}
Thus the equal phase paths connect to the singular points \eqref{eqn:sing} with separation angles given by
\begin{align}
\sigma-\frac{m}{2} \to \frac{n\pi}{2}, \quad & \text{as} \quad \rho\to 0.
\end{align}
Moreover, as $\rho\to0$ we have that 
\begin{align}
\Re(K)\to -\infty \quad&\text{if}\quad -\frac{3\pi}{4}<\sigma-\frac{m}{2}<-\frac{\pi}{4}, \\
\Re(K)\to \infty \quad&\text{if}\quad -\frac{3\pi}{4}<\sigma-\frac{m}{2}<-\frac{\pi}{4}.
\end{align}
Thus the equal-phase contours connect to the singular points at intervals of $\pi/2$ and are located alternatingly in the centre of the hills and valleys of $\Re(K)$ (see Fig.~\ref{fig:deformation}. For every choice of $\theta$, the integration path $[0,2\pi]$ enters the singular points in a valley of $\Re(K)$ (and thus the integral is well defined). For a given valley, deformation that complies with Cauchy's theorem must preserve for each valley the number of paths entering minus those exiting throughout the deformation. In region $\bm{B}$ (see Fig.~\ref{fig:regions}) it is not possible to deform in such a way that also produce a steepest descent path which passes through transverse type saddles. 

\end{document}